\documentclass[11pt]{article}

\usepackage[margin=1in]{geometry}

\usepackage{graphics}
\usepackage{epsfig}
\usepackage{epstopdf}
\usepackage{amsmath}
\usepackage{array}
\begin{document}

\begin{center}
{\LARGE Bulk viscous matter and recent acceleration of the Universe. \\[0.2in]}
{ Athira Sasidharan and Titus K Mathew\\
e-mail:athirasnair91@cusat.ac.in, titus@cusat.ac.in \\ Department of
Physics, Cochin University of Science and Technology, Kochi, India.}
\end{center}

\begin{abstract}
We consider a cosmological model dominated by bulk viscous matter
with total bulk viscosity coefficient proportional to the velocity
and acceleration of the expansion of the universe in such a way that
$\zeta=\zeta_{0}+\zeta_{1}\frac{\dot{a}}{a}+\zeta_{2}\frac{\ddot{a}}{\dot{a}}.$
We show that there exist two limiting conditions in the bulk viscous
coefficients, ($\zeta_{0}$, $\zeta_{1}$, $\zeta_{2}$) which
corresponds to a universe having a Big-Bang at the origin, followed
by an early decelerated epoch and then making a smooth transition
into an accelerating epoch. We have constrained the model using the
type Ia Supernovae data, evaluated the best estimated values of all
the bulk viscous parameters and the Hubble parameter corresponding
to the two limiting conditions. We found that even though the
evolution of the cosmological parameters are in general different
for the two limiting cases, they show identical behavior for the
best estimated values of the parameters from both the limiting
conditions. A recent acceleration would occur if
$\tilde{\zeta}_{0}+\tilde{\zeta}_{1}>1$ for the first limiting
conditions and if $\tilde{\zeta}_{0}+\tilde{\zeta}_{1}<1$ for the
second limiting conditions. The age of the universe predicted by
this model is found to be less than that predicted from the oldest
galactic globular clusters. The total bulk viscosity seems to be
negative in the past and becomes positive when $z\leq0.8$. So the
model violates the local second law of thermodynamics. However, the
model satisfies the generalized second law of thermodynamics at the
apparent horizon throughout the evolution of the universe. We also
made a statefinder analysis of the model and found that it is
distinguishably different from the standard $\Lambda$CDM model at
present, but shows a de Sitter type behavior in the far future of
the evolution.
\end{abstract}

\section{Introduction}
\label{intro} Observational data on type Ia Supernovae have shown
that the current universe is accelerating and the acceleration began
in the recent past of the universe \cite{Riess1,Perl1}. This was
further supported by the observations on cosmic microwave background
radiations (CMBR) \cite{Bennet1} and large scale structure
\cite{Tegmark1}. Despite the mounting observational evidence on this
recent acceleration, its nature and fundamental origin is still an
open question. Many models has been proposed to explain this current
acceleration. Basically there are two approaches. The first one is
to modify the right hand side of the Einstein's equation by
considering specific forms for the energy-momentum tensor $T_{\mu
\nu }$ having a negative pressure, which culminate in the proposal
of an exotic energy called dark energy. The simplest candidate for
dark energy is the so-called cosmological constant $\Lambda ,$\
which is characterized by the equation of state, $\omega_{\Lambda}=
-1$ and a constant energy density \cite{weinberg1}. However, it
faced with many drawbacks. Of these, the two main problems are the
coincidence problem and the fine tuning problem \cite{cope}.
Coincidence problem refers to the coincidence of densities of dark
matter and dark energy, even though their evolutions are different,
and the fine tuning problem refers to the discrepancy between the
theoretical and the observational value of the vacuum constant or
cosmological constant, which is assumed to drive the accelerated
expansion. These discrepancies motivated the consideration of
various dynamical dark energy models like quintessence
\cite{fujii,carroll}, k-essence \cite{chiba1} and perfect fluid
models (like Chaplygin gas model) \cite{kamen1}. The second approach
for explaining the current acceleration of the universe is to modify
the left hand side of the Einstein's equation, i.e., the geometry of
the space time. The models that belong to this class (modified
gravity) are the so called $f(R)$ gravity \cite{capo1}, $f(T)$\
gravity \cite{ferraro1}, Gauss-Bonnet theory \cite{nojiri}, Lovelock
gravity \cite{pad2}, Horava-Lifshitz gravity \cite{horava1},
scalar-tensor theories \cite{amendola1}, braneworld models
\cite{dvali1} etc.

It was noted by several authors that the bulk viscous fluid can
produce acceleration in the expansion of the universe. This was
first studied in the context of inflationary phase in the early
universe \cite{pad,waga}. In the context of late acceleration of the
universe, the effect of bulk viscous fluid was studied in references
\cite{fabris1,li1,Hiplito1,av1,av2}. But a shortcoming in
considering the bulk viscous fluid is the problem of finding out a
viable mechanism for the origin of bulk viscosity in the expanding
universe. From the theoretical point of view, bulk viscosity can
arise due to deviations from the local thermodynamic equilibrium
\cite{Zimdahl1}. In cosmology, bulk viscosity arises as an effective
pressure to restore the system back to its thermal equilibrium,
which was broken when the cosmological fluid expands (or contract)
too fast. This bulk viscosity pressure generated, ceases as soon as
the fluid reaches the thermal equilibrium
\cite{wilson1,okumura1,ilg1}.

In this paper, we analyze matter dominated cosmological model with
bulk viscosity with reference to the question whether it can cause
the recent acceleration of the universe. We took the bulk viscosity
coefficient as proportional to both the velocity and acceleration of
the expansion of the universe. The matter is basically a
pressureless fluid comprising both baryonic and dark matter
components. If the bulk viscous matter produce the recent
acceleration of the universe then it would unify the description of
both dark matter and dark energy. The advantage is that it
automatically solves the coincidence problem because there is no
separate dark energy component. A similar model was studied by
Avelino et al. \cite{av3}, but in constraining the parameters,
($\zeta_{0}$, $\zeta_{1}$, $\zeta_{2}$) using the observational data
the authors fixed either $\zeta_{1}$ or $\zeta_{2}$ as zero. So it
is effectively a two parameter model. In this reference the authors
have ruled out the possibility of bulk viscous matter to be a
candidate for dark energy. We think that one should study the model
by evaluating all the parameters simultaneously, which may lead to a
more mature conclusion regarding the status of bulk viscous dark
matter as dark energy. In the present work we aim to such an
analysis in studying the evolution of all the cosmological
parameters by simultaneously evaluating all the constant parameters
on which the total bulk viscous coefficient depends.

The paper is organized as follows. In section \ref{sec:2} we present
the basic formalism of the bulk viscous matter dominated flat
universe. We derive the Hubble parameter in this section. In section
\ref{sec:3}, we identify two different limiting conditions for the
bulk viscous coefficients corresponding to which the universe begins
with a Big-Bang, followed by an early decelerated epoch and then
entering a phase of recent acceleration. We also present the
evolution of the scale factor and age of the universe in this
section. In section \ref{sec:4a} we study the evolution of the
cosmological parameters like deceleration parameter, the equation of
state parameter, matter density and curvature scalar. Section
\ref{sec:8} consists of the study of the status of local second law
and generalized second law of thermodynamics in the model. In
section \ref{sec:9} we presents the statefinder analysis of the
model to contrast it with other standard models of dark energy. The
estimation of parameters using type Ia Supernova data is given in
section \ref{sec:estmation}, followed by conclusions in section
\ref{sec:conclu}.

\section{FLRW Universe dominated with bulk viscous matter}
\label{sec:2} We consider a spatially flat universe described by the
Friedmann-Lemaitre-Robertson-Walker (FLRW) metric,
\begin{equation}
ds^{2}=-dt^{2}+a(t)^{2}(dr^{2}+r^{2}d\theta ^{2}+r^{2}\sin
^{2}\theta d\phi ^{2})
\end{equation}
where $(r,\theta,\phi)$ are the co-moving coordinates, $t$ is the
cosmic time and $a(t)$ is the scale factor of the universe dominated
with bulk viscous matter, which can produce an effective pressure
\cite{weinberg2,misner1}
\begin{equation}
\label{p} P^{*}=P-3\zeta H
\end{equation}
where $P$ is the normal pressure, which is zero for non-relativistic
matter and $\zeta$ is the coefficient of bulk viscosity, which can
be a function of Hubble Parameter and its derivatives in an
expanding universe. We have not considered the radiation component,
as it is a reasonable simplification as long as we are concerned
with late time acceleration. The form of equation (\ref{p}) was
originally proposed by Eckart in 1940 \cite{Eckart1}. A similar
theory was also proposed by Landau and Lifshitz \cite{Landau1}.
However, Eckart theory suffer from pathologies. One of them is that
in this theory, dissipative perturbations propagate at infinite
speeds \cite{Israel1}. Another one is that the equilibrium states in
the theory are unstable \cite{Hiscock1}. In 1979, Israel and Stewart
\cite{Israel2,Israel3} developed a more general theory which was
causal and stable and one can obtain the Eckart theory from it in
the first order limit, when the relaxation time goes to zero. So, in
the limit of vanishing relaxation time, the Eckart theory is a good
approximation to the Israel-Stewart theory.

Even though Eckart theory have drawbacks, it is less complicated
than the Israel-Stewart theory. So it has been used widely by many
authors to characterize the bulk viscous fluid. For example in
references \cite{fabris1,kremer1,cataldo1,hu1}, Eckart approach has
been used in dealing with the accelerating universe with the bulk
viscous fluid. In this context, it is reasonable to point out that
Hiscock et. al.\cite{hiscock1} have found that pathological Eckart
theory and also truncated Israel- Stewart theory (avoiding the
non-linear terms) can produce early inflation. However, as pointed
out by the same authors, in the truncated version of Israel-Stewart
theory, the relaxation time stands to be a constant which is in fact
not logically correct in an expanding universe. However, there exist
some later studies \cite{Zimdahl,Zakari} which deals with the
importance of equation of state in such theories inorder to explain
the acceleration. But, it should be checked whether these theories
will produce the late acceleration of the universe as observed
today. One should also note at this juncture that a more general
formulation than Israel-Stewart model was proposed by Pavon et al.
\cite{pavon1} for irreversible process, especially in dealing with
thermodynamic equilibrium of dissipative fluid.

The Friedmann equations describing the evolution of flat universe
dominated with bulk viscous matter are,
\begin{equation}
\label{friedmann} H^{2}=\frac{\rho_{m}}{3}
\end{equation}
\begin{equation}
2\frac{\ddot{a}}{a}+\left(\frac{\dot{a}}{a}\right)^{2}=-P^{*}
\end{equation}
where we have taken $8\pi G = 1$, $\rho_{m}$ is the matter density
and overdot represents the derivative with respect to cosmic time
$t$. The conservation equation is
\begin{equation}
\label{conser} \dot{\rho}_{m}+3H(\rho_{m}+P^{*})=0.
\end{equation}

In this paper we consider a parameterized bulk viscosity of the form
\cite{ren1},
\begin{equation}
\label{zeta}
\zeta=\zeta_{0}+\zeta_{1}\frac{\dot{a}}{a}+\zeta_{2}\frac{\ddot{a}}{\dot{a}}.
\end{equation}
 In an expanding universe, the bulk viscosity coefficient may
depends on both the velocity and acceleration. The most logical form
can be a linear combination of three terms: the first term is a
constant $\zeta_{0}$, the second term is proportional to the Hubble
parameter, which characterizes the dependence of the bulk viscosity
on velocity, and the third is proportional to
$\frac{\ddot{a}}{\dot{a}}$, characterizing the effect of
acceleration on the bulk viscosity. Moreover, such a form for the
bulk viscous coefficient implies the most general form of the
equation of state \cite{ren1}. In terms of Hubble parameter
$H=\frac{\dot{a}}{a}$, this can be written as,
\begin{equation}
\label{z} \zeta=\zeta_{0}+\zeta_{1}H+\zeta_{2}(\frac{\dot{H}}{H}+H)
\end{equation}
From Friedmann equations, and from equations (\ref{p}),
(\ref{conser}) and (\ref{z}), we can obtain a first order
differential equation for Hubble parameter by replacing
$\frac{d}{dt}$ with $\frac{d}{d\ln a}$ through
$\frac{d}{dt}=H\frac{d}{d\ln a}$ as,

\begin{equation}
\label{derihubble} \frac{dH}{d\ln
a}-\left(\frac{\tilde{\zeta}_{1}+\tilde{\zeta}_{2}-3}{2-\tilde{\zeta}_{2}}\right)H-\left(\frac{\tilde{\zeta}_{0}}{2-\tilde{\zeta}_{2}}\right)H_{0}=0
\end{equation}
where
\begin{equation}
\label{dimen} \tilde{\zeta}_{0}=\frac{3\zeta_{0}}{H_{0}},\ \ \ \
\tilde{\zeta}_{1}=3\zeta_{1},\ \ \ \ \tilde{\zeta}_{2}=3\zeta_{2}
\end{equation}
are the dimensionless bulk viscous parameters and $H_{0}$ is the
present value of the Hubble parameter. The above equation can be
integrated to obtain the Hubble parameter as,

\begin{equation}
\label{hubbleina}
H(a)=H_{0}\left[a^{\frac{\tilde{\zeta}_{1}+\tilde{\zeta}_{2}-3}{2-\tilde{\zeta}_{2}}}\left(1+\frac{\tilde{\zeta}_{0}}{\tilde{\zeta}_{1}+\tilde{\zeta}_{2}-3}\right)-\frac{\tilde{\zeta}_{0}}{\tilde{\zeta}_{1}+\tilde{\zeta}_{2}-3}\right]
\end{equation}
This equation shows that when $\tilde{\zeta}_{0}$,
$\tilde{\zeta}_{1}$ and $\tilde{\zeta}_{2}$ are all zeros, the
Hubble parameter, $H=H_0a^{-\frac{3}{2}}$ which corresponds to the
ordinary matter dominated universe.
 When
$\tilde{\zeta}_{1}=\tilde{\zeta}_{2}=0$, the Hubble parameter
reduces to \cite{av1}
\begin{equation}
H(a)=H_{0}\left[a^{-\frac{3}{2}}\left(1-\frac{\tilde{\zeta}_{0}}{3}\right)+\frac{\tilde{\zeta}_{0}}{3}\right].
\end{equation}
\section{Behavior of scale factor and age of the universe}
\label{sec:3} 
\begin{figure}
\centering
\includegraphics[scale=0.6]{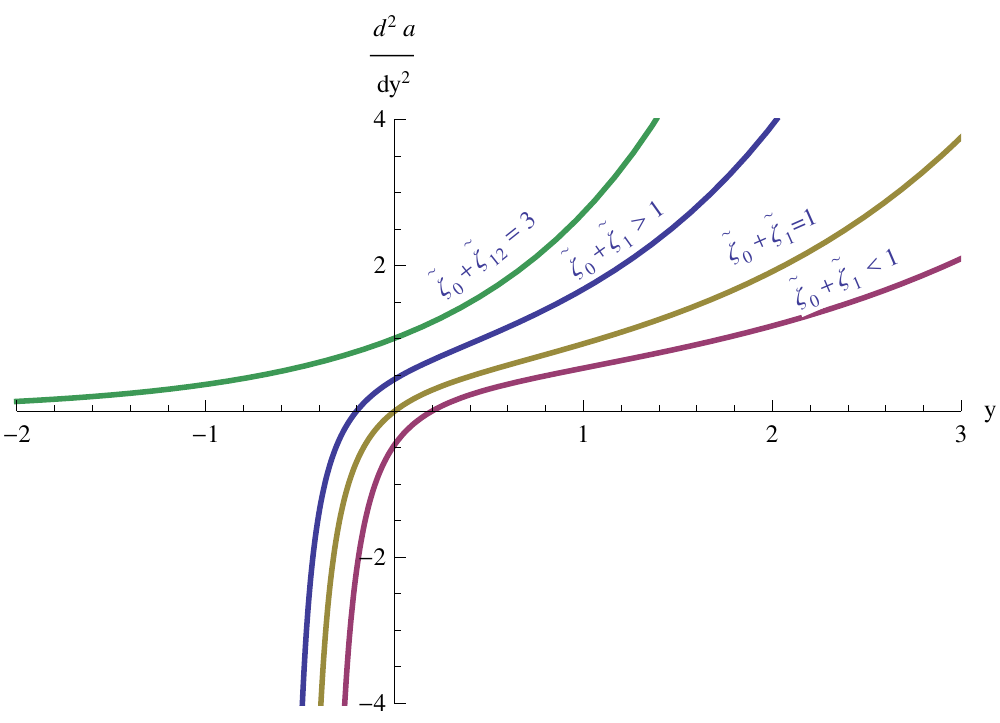}
\caption{\label{fig:secder1} Evolution of the second derivative of
the scale factor with respect to $y=H_{0}(t-t_{0})$ for the first
limiting conditions of parameters, $\tilde{\zeta}_{0}>0, \,
\tilde{\zeta}_{0}+\tilde{\zeta}_{12}<3, \, \tilde{\zeta}_{12}<3, \,
\tilde{\zeta}_{2}<2$. The curve corresponding to
$\tilde{\zeta}_{0}+\tilde{\zeta}_{12}\geq3$ represents a universe
which is eternally accelerating. If
$\tilde{\zeta}_{0}+\tilde{\zeta}_{1}>1$, the transition to the
accelerating epoch happens in the past.If
$\tilde{\zeta}_{0}+\tilde{\zeta}_{1}<1$ the transition will be in
the future. If $\tilde{\zeta}_{0}+\tilde{\zeta}_{1}=1$, the
transition occurs at present.}
\end{figure}

\begin{figure}
\centering
\includegraphics[scale=0.6]{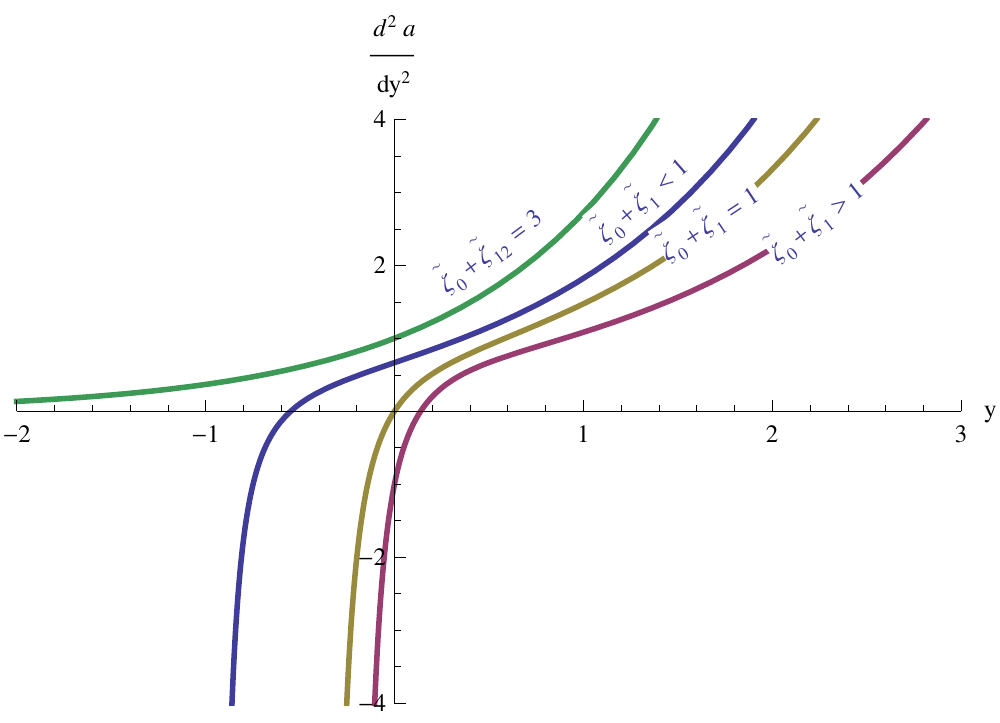}
\caption{\label{fig:secder2} Evolution of the second derivative of
the scale factor with respect to $y=H_{0}(t-t_{0})$ for the second
limiting conditions of parameters, $\tilde{\zeta}_{0}<0, \,
\tilde{\zeta}_{0}+\tilde{\zeta}_{12}>3, \, \tilde{\zeta}_{12}>3, \,
\tilde{\zeta}_{2}>2$. The curve corresponding to
$\tilde{\zeta}_{0}+\tilde{\zeta}_{12}\leq3$ represents a universe
which is eternally accelerating. If
$\tilde{\zeta}_{0}+\tilde{\zeta}_{1}<1$, the transition to the
accelerating epoch happens in the past.If
$\tilde{\zeta}_{0}+\tilde{\zeta}_{1}>1$ the transition will be in
the future. If $\tilde{\zeta}_{0}+\tilde{\zeta}_{1}=1$, the
transition occurs at present.}
\end{figure}

In this section we analyze the behavior of scale
factor in a bulk viscous matter dominated universe. Using the
definition of Hubble parameter, equation (\ref{hubbleina}) becomes,
\begin{equation}
\frac{1}{a}\frac{da}{dt}=H_{0}\left[a^{\frac{\tilde{\zeta}_{12}-3}{2-\tilde{\zeta}_{2}}}\left(1+\frac{\tilde{\zeta}_{0}}{\tilde{\zeta}_{12}-3}\right)-\frac{\tilde{\zeta}_{0}}{\tilde{\zeta}_{12}-3}\right]
\end{equation}
where $\tilde{\zeta}_{12}=\tilde{\zeta}_{1}+\tilde{\zeta}_{2}$.
Integrating the above equation to solve for the scale factor we get,

\begin{equation}
\label{scalefactor}
a(t)=\left[(\frac{\tilde{\zeta}_{0}+\tilde{\zeta}_{12}-3}{\tilde{\zeta}_{0}})+(\frac{3-\tilde{\zeta}_{12}}{\tilde{\zeta}_{0}})
e^{\frac{\tilde{\zeta}_{0}}{2-\tilde{\zeta}_{2}}H_{0}(t-t_{0})}\right]^{\frac{2-\tilde{\zeta}_{2}}{3-\tilde{\zeta}_{12}}}
\end{equation}

where $t_0$ is the present cosmic time. Assuming, $y=H_{0}(t-t_{0})$
and taking second derivative of the scale factor $a$ (equation
(\ref{scalefactor})) with respect to $y$, we obtain

\begin{equation}
\begin{aligned}
\begin{split}
\label{seconda}
\frac{d^{2}a}{dy^{2}}=&\frac{e^{\frac{\tilde{\zeta}_{0}y}{2-\tilde{\zeta}_{2}}}}{2-\tilde{\zeta}_{2}}\left[\tilde{\zeta}_{0}+\tilde{\zeta}_{12}-3+(2-\tilde{\zeta}_{2})e^{\frac{\tilde{\zeta}_{0}y}{2-\tilde{\zeta}_{2}}}\right]
\\ & \left[\frac{\tilde{\zeta}_{0}+\tilde{\zeta}_{12}-3+(3-\tilde{\zeta}_{12})e^{\frac{\tilde{\zeta}_{0}y}{2-\tilde{\zeta}_{2}}}}{\tilde{\zeta}_{0}}\right]^{\frac{2(\tilde{\zeta}_{1}-2)+\tilde{\zeta}_{2}}{3-\tilde{\zeta}_{12}}}.
\end{split}
\end{aligned}
\end{equation}
From the behavior of the scale factor and the Hubble parameter, it
is possible to identify two limiting conditions on
$(\tilde{\zeta}_{0},\tilde{\zeta}_{1},\tilde{\zeta}_{2})$ which
corresponds to a universe that would start with a Big-Bang followed
by an early decelerated epoch, then making a transition into the
accelerated epoch in the later times. These two conditions are,
\begin{equation}
\label{condition1}
 \tilde{\zeta}_{0} > 0, \,  \tilde{\zeta}_{12} < 3, \, \tilde{\zeta}_{2} < 2
\end{equation}
\begin{equation}
\label{condition2} \tilde{\zeta}_{0} < 0, \,  \tilde{\zeta}_{12} >
3, \, \tilde{\zeta}_{2} > 2.
\end{equation}
The first condition is to be simultaneously satisfied with
$\tilde{\zeta}_{0}+\tilde{\zeta}_{12}<3$ and the second condition
with  $\tilde{\zeta}_{0}+\tilde{\zeta}_{12}>3$. Instead of these, if
the first condition (\ref{condition1}) is satisfied simultaneously
with $\tilde{\zeta}_{0}+\tilde{\zeta}_{12}>3$ or the second
condition (\ref{condition2}) with
$\tilde{\zeta}_{0}+\tilde{\zeta}_{12}<3$, then the universe will
undergo an eternally accelerated expansion, see the curve for
$\tilde{\zeta}_{0}+\tilde{\zeta}_{12}=3$ in figures
\ref{fig:secder1} and \ref{fig:secder2}. We have obtained the best
estimates of the bulk viscous parameters
$(\tilde{\zeta}_{0},\tilde{\zeta}_{1},\tilde{\zeta}_{2})$
corresponding to the cases, equations (\ref{condition1}) and
(\ref{condition2}) separately, using the SCP ``Union" SNe Ia data
set, about which we will discuss in section \ref{sec:estmation}.

\begin{figure}
\centering
\includegraphics[scale=0.6]{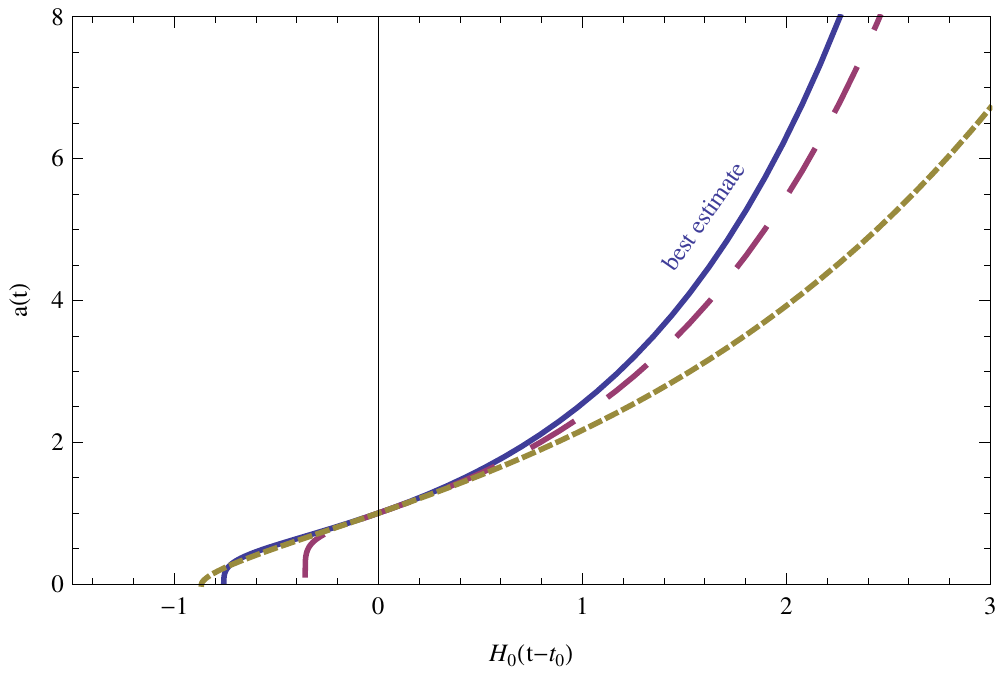}
\caption{\label{fig:scalefactor1} Behavior of the scale factor for
the first limiting conditions of parameters, $\tilde{\zeta}_{0}>0$,
$\tilde{\zeta}_{0}+\tilde{\zeta}_{12}<3$, $\tilde{\zeta}_{12}<3$,
$\tilde{\zeta}_{2}<2$. Solid line corresponds to the best fit
parameters
$(\tilde{\zeta}_{0},\tilde{\zeta}_{1},\tilde{\zeta}_{2})=(7.83,-5.13,-0.51)$.
Dashed line corresponds to parameter values $(5, -4, 1)$ and the
dotted line corresponds to values $(4, -2, -3)$. The parameter
values are selected so that the transition to the accelerated epoch
happens in the past.}
\end{figure}

\begin{figure}
\centering
\includegraphics[scale=0.6]{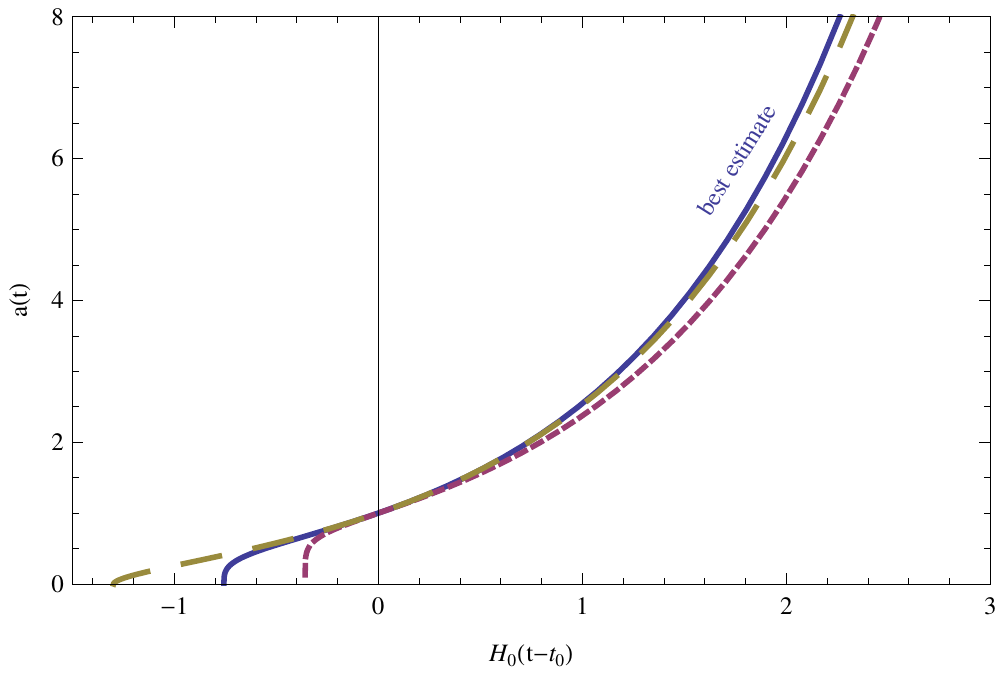}
\caption{\label{fig:scalefactor2} Behavior of the scale factor for
the second limiting conditions of parameters, $\tilde{\zeta}_{0}<0$,
$\tilde{\zeta}_{0}+\tilde{\zeta}_{12}>3$, $\tilde{\zeta}_{12}>3$,
$\tilde{\zeta}_{2}>2$. Solid line corresponds to the best fit
parameters
$(\tilde{\zeta}_{0},\tilde{\zeta}_{1},\tilde{\zeta}_{2})=(-4.68,
4.67, 3.49)$. Dashed line corresponds to parameter values $(-6, 4,
6)$ and the dotted line corresponds to values $(-5, 6, 3)$. The
parameter values are selected so that the transition to the
accelerated epoch happens in the past.}
\end{figure}

For both the cases of bulk viscous parameters, as given by equations
(\ref{condition1}) and (\ref{condition2}), the Hubble parameter
given in equation (\ref{hubbleina}) becomes infinity as the scale
factor $a\to 0$, which implies that the density becomes infinity at
the origin, indicating the presence of a Big-Bang at the origin. The
behavior of the scale factor as given in equation
(\ref{scalefactor}) are shown in figures \ref{fig:scalefactor1} and
\ref{fig:scalefactor2} for the two conditions of parameters
respectively. As $(t-t_{0})\rightarrow0$, the scale factor in both
the cases reduces to
\begin{equation}
a(t)\to\left[1+\frac{3-\tilde{\zeta}_{12}}{2-\tilde{\zeta}_{2}}H_{0}(t-t_{0})\right]^{\frac{2-\tilde{\zeta}_{2}}{3-\tilde{\zeta}_{12}}},
\end{equation}
which corresponds to an early decelerated expansion. In both the
cases of limiting conditions, as $(t-t_{0})\rightarrow\infty$, the
scale factor tends to,
\begin{equation}
a(t)\to
e^{\frac{\tilde{\zeta}_{0}}{2-\tilde{\zeta}_{2}}H_{0}(t-t_{0})}.
\end{equation}
This corresponds to acceleration similar to the de Sitter phase
which implies that the bulk viscous dark matter behaves similar to
the cosmological constant as $(t-t_{0})\rightarrow\infty$, at least
at the background level. An important point to be noted is that the
evolution of the scale factor is the same for the best estimates of
the bulk viscous coefficient from the two limiting conditions, see
figures \ref{fig:scalefactor1} and \ref{fig:scalefactor2}.

The scale factor and red shift corresponding to the transition from
decelerated to accelerated expansion can be obtained as shown below.
From the Hubble parameter (equation (\ref{hubbleina})) the
derivative of $\dot{a}$ with respect to $a$ can be obtained as,
\begin{equation}
\label{transeq}
\frac{d\dot{a}}{da}=H_{0}\left[\left(\frac{\tilde{\zeta}_{1}-1}{2-\tilde{\zeta}_{2}}\right)\left(\frac{\tilde{\zeta}_{0}+\tilde{\zeta}_{12}-3}{\tilde{\zeta}_{12}-3}\right)a^{\frac{\tilde{\zeta}_{12}-3}{2-\tilde{\zeta}_{2}}}-\frac{\tilde{\zeta}_{0}}{\tilde{\zeta}_{12}-3}\right].
\end{equation}
Equating this to zero, we obtain the transition scale factor
$a_{T}$,
\begin{equation}
\label{transa}
a_{T}=\left[\frac{\tilde{\zeta}_{0}\left(2-\tilde{\zeta}_{2}\right)}{\left(\tilde{\zeta}_{1}-1\right)\left(\tilde{\zeta}_{0}+\tilde{\zeta}_{12}-3\right)}\right]^{\frac{2-\tilde{\zeta}_{2}}{\tilde{\zeta}_{12}-3}}
\end{equation}
and the corresponding transition red shift $z_{T}$ is,
\begin{equation}
\label{transz}
z_{T}=\left[\frac{\tilde{\zeta}_{0}\left(2-\tilde{\zeta}_{2}\right)}{\left(\tilde{\zeta}_{1}-1\right)\left(\tilde{\zeta}_{0}+\tilde{\zeta}_{12}-3\right)}\right]^{-{\frac{2-\tilde{\zeta}_{2}}{\tilde{\zeta}_{12}-3}}}-1.
\end{equation}
From equations (\ref{transa}) and (\ref{transz}), it is clear that
if $\tilde{\zeta}_{0}+\tilde{\zeta}_{1}=1$, the transition from
decelerated phase to accelerated phase occurs at $a_{T}=1$ and
$z_{T}=0$, which corresponds to the present time of the universe.
For the first case of limiting conditions of parameters with
$\tilde{\zeta}_{0}>0$, the transition would takes place in the past
if $\tilde{\zeta}_{0}+\tilde{\zeta}_{1}>1$ and in the future if
$\tilde{\zeta}_{0}+\tilde{\zeta}_{1}<1$. For the second case of
limiting conditions of parameters, that corresponds to
$\tilde{\zeta}_{0}<0$, the above conditions are reversed such that
transition would takes place in the future if
$\tilde{\zeta}_{0}+\tilde{\zeta}_{1}>1$ and in the past if
$\tilde{\zeta}_{0}+\tilde{\zeta}_{1}<1$. These are shown in figures
\ref{fig:secder1} and \ref{fig:secder2} respectively, where we have
plotted $\frac{d^{2}a}{dy^{2}}$ (equation \ref{seconda}) with $y$.

Age of the universe can be deduced from the scale factor equation
(\ref{scalefactor}) by equating it to zero. The time elapsed since
the Big-Bang is,
\begin{equation}
t_{B}=
t_{0}+\left(\frac{2-\tilde{\zeta}_{2}}{H_{0}\tilde{\zeta}_{0}}\right)\ln\left(1-\frac{\tilde{\zeta}_{0}}{3-\tilde{\zeta}_{12}}\right).
\end{equation}
Hence, the age of the universe since Big-Bang is
\begin{equation}
Age\equiv
t_{0}-t_{B}=-\left(\frac{2-\tilde{\zeta}_{2}}{H_{0}\tilde{\zeta}_{0}}\right)ln\left(1-\frac{\tilde{\zeta}_{0}}{3-\tilde{\zeta}_{12}}\right).
\end{equation}

\begin{figure}
\centering
\includegraphics[scale=0.6]{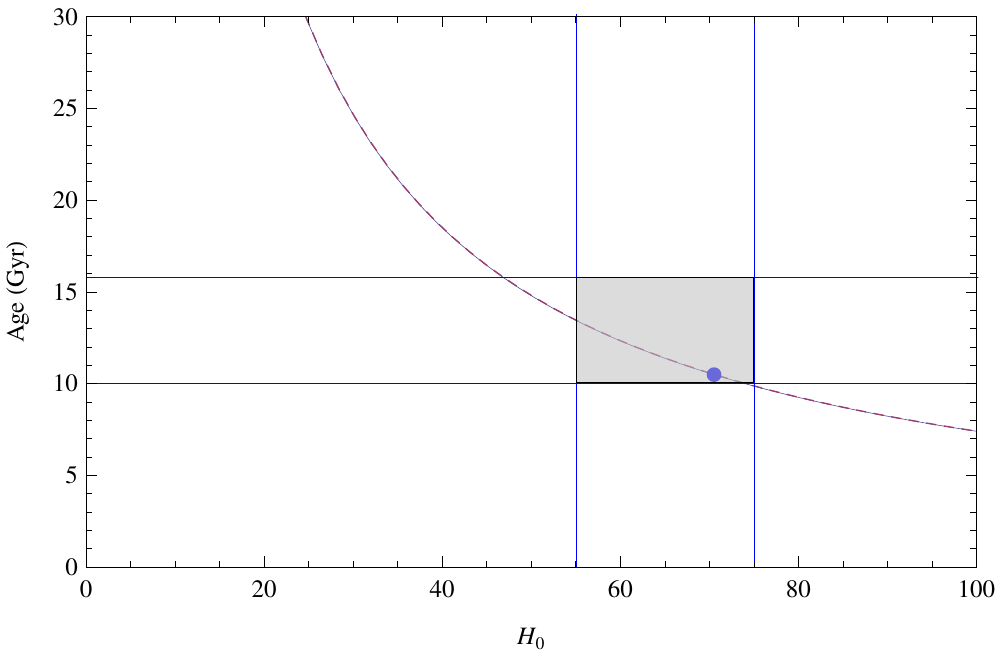}
\caption{\label{fig:age} Plot of the age of the universe in Gyr with
$H_{0}$ in units of $km s^{-1}Mpc^{-1}$ for the best fit values of
the bulk viscous parameters. The plots are identical for the best
estimated values of the parameters from both the limiting
conditions. The point located in the figure corresponds to an age
10.5 Gyr for the best estimate value of $H_0$, obtained as $70.49 km
s^{-1}Mpc^{-1}$. The shaded region corresponds to the interval $H_0
(55,75)km s^{-1}Mpc^{-1}$ and age $(10,15.8)$ Gyr, which are the
permitted intervals for $H_{0}$ and age, derived using observations
on Galactic globular clusters from the Hipparcos parallaxes
\cite{carretta1}.}
\end{figure}

A plot of the age of the universe with $H_{0}$ for the best
estimates of the bulk viscous parameters is shown in figure
\ref{fig:age} (the evolution is the same for the best estimates from
the two limiting conditions). The age of the universe corresponding
to the best estimates of the present Hubble parameter is found to be
$10.90$ Gyr and is marked in the plot. This value is less compared
to the age deduced from CMB anisotropy data \cite{tegmark2} and also
that from the oldest globular clusters \cite{carretta1}, which is
around $12.9\pm2.9$ Gyr. For comparison, we have also extracted the
value of the Hubble parameter for the $\Lambda$CDM model using the
same data set (see Table \ref{tab:1} in section \ref{sec:estmation})
from which the age of the universe is found to be around 13.85 Gyr.
So compared to the age of the universe from globular clusters and
the standard $\Lambda$CDM model, the present model, where the bulk
viscous matter replaces the dark energy, predicts relatively a low
age.
\section{Cosmological parameters}
\label{sec:4a}
\subsection{Deceleration parameter} \label{sec:4} The results
regarding the transition of the universe into the accelerated epoch
discussed in the above section can be further verified by studying
the evolution of the deceleration parameter $q,$ which is defined
as,
\begin{equation}
q(a)=-\frac{\ddot{a}a}{\dot{a}^{2}}=-\frac{\ddot{a}}{a}\frac{1}{H^{2}}.
\end{equation}
For the bulk viscous matter dominated universe, one can obtain using
Friedmann equations,
\begin{equation}
\label{adouble}
\frac{\ddot{a}}{a}=-\frac{1}{6}\left[\rho_{m}-9H\left(\zeta_{0}+\zeta_{1}H+\zeta_{2}\left(\frac{\dot{H}}{H}+H\right)\right)\right].
\end{equation}
Using the dimensionless bulk viscous parameters as defined in
equation (\ref{dimen}) and using equations (\ref{friedmann}) and
(\ref{adouble}), the deceleration parameter becomes,
\begin{equation}
q=\frac{1}{2}\left[1-\left(\frac{H_{0}}{H}\tilde{\zeta}_{0}+\tilde{\zeta}_{1}+\frac{\tilde{\zeta}_{2}}{H}\left(\frac{\dot{H}}{H}+H\right)\right)\right].
\end{equation}
Substituting equations(\ref{derihubble}) and (\ref{hubbleina}), we
can obtain the deceleration parameter in terms of $a$,
$\tilde{\zeta}_{0}$, $\tilde{\zeta}_{1}$ and $\tilde{\zeta}_{2}$ as,
\begin{equation}
\label{decelerationa}
q(a)=\frac{1}{2-\tilde{\zeta}_{2}}\left[1-\tilde{\zeta}_{1}-\frac{\tilde{\zeta}_{0}}{a^{\frac{\tilde{\zeta}_{12}-3}{2-\tilde{\zeta}_{2}}}\left[1+\frac{\tilde{\zeta}_{0}}{\tilde{\zeta}_{12}-3}\right]-\frac{\tilde{\zeta}_{0}}{\tilde{\zeta}_{12}-3}}\right].
\end{equation}
In terms of red shift, the above equation becomes,
\begin{equation}
\label{deceleration}
q(z)=\frac{1}{2-\tilde{\zeta}_{2}}[1-\tilde{\zeta}_{1}-\frac{\tilde{\zeta}_{0}}{(1+z)^{-\frac{\tilde{\zeta}_{12}-3}{2-\tilde{\zeta}_{2}}}[1+\frac{\tilde{\zeta}_{0}}{\tilde{\zeta}_{12}-3}]-\frac{\tilde{\zeta}_{0}}{\tilde{\zeta}_{12}-3}}].
\end{equation}

\begin{figure}
\centering
\includegraphics[scale=0.6]{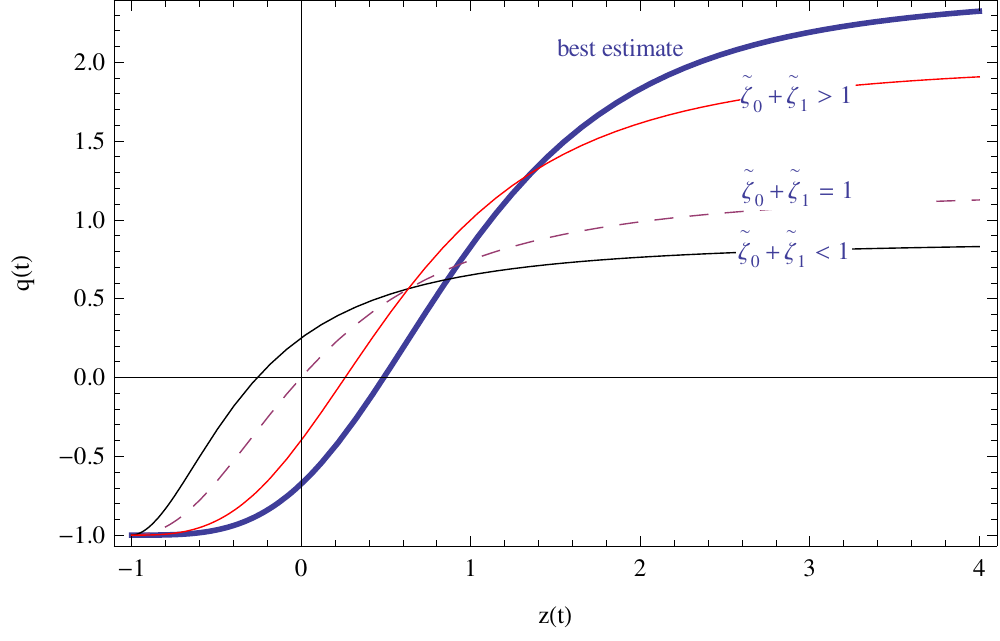}
\caption{\label{fig:deceleration1} Evolution of the deceleration
parameter with red shift for the first limiting conditions of
viscous parameters, $\tilde{\zeta}_{0}>0, \,
\tilde{\zeta}_{0}+\tilde{\zeta}_{12}<3, \, \tilde{\zeta}_{12}<3, \,
\tilde{\zeta}_{2}<2$. $q$ enters the negative region in the recent
past if $\tilde{\zeta}_{0}+\tilde{\zeta}_{1}>1$, at present if
$\tilde{\zeta}_{0}+\tilde{\zeta}_{1}=1$ and in the future if
$\tilde{\zeta}_{0}+\tilde{\zeta}_{1}<1.$ Evolution of $q$ for the
best estimated values of the bulk viscous parameters is also shown.
The redshift at which the $q$ enters the negative region for the
best estimated values of the bulk viscous parameters corresponds to
$z_T=0.49^{+0.075}_{-0.057}.$}
\end{figure}
\begin{figure}
\centering
\includegraphics[scale=0.6]{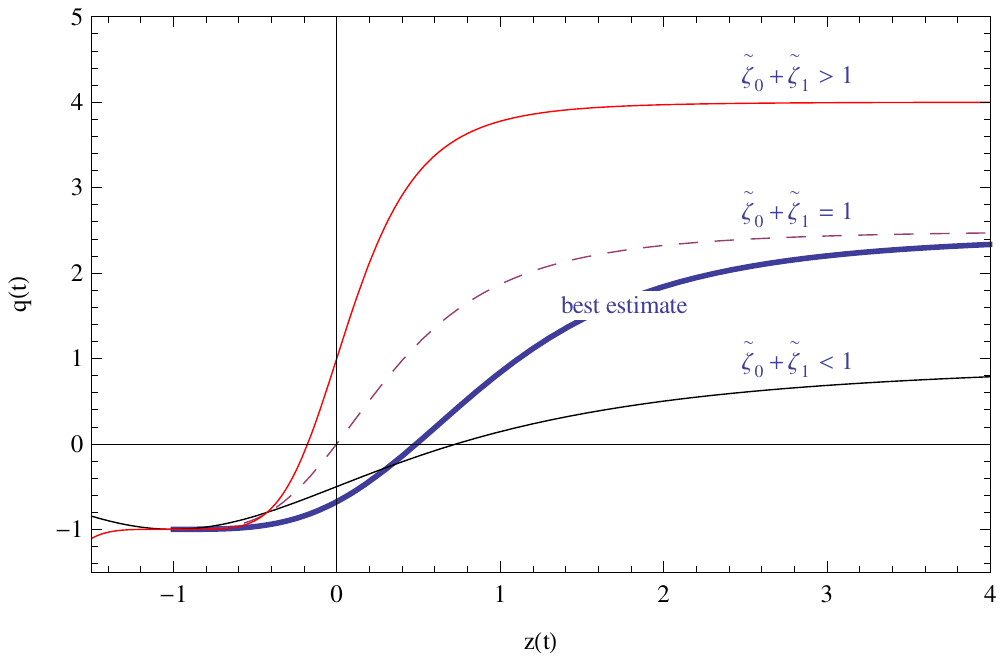}
\caption{\label{fig:deceleration2} Evolution of the deceleration
parameter with red shift for the second limiting conditions of
viscous parameters, $\tilde{\zeta}_{0}<0, \,
\tilde{\zeta}_{0}+\tilde{\zeta}_{12}>3, \, \tilde{\zeta}_{12}>3, \,
\tilde{\zeta}_{2}>2.$ $q$ enters the negative region in the recent
past if $\tilde{\zeta}_{0}+\tilde{\zeta}_{1}<1$, at present if
$\tilde{\zeta}_{0}+\tilde{\zeta}_{1}=1$ and in the future if
$\tilde{\zeta}_{0}+\tilde{\zeta}_{1}>1.$ Evolution of $q$ for the
best estimated values of the bulk viscous parameters is also shown.
The redshift at which the $q$ enters the negative region for the
best estimated values of the bulk viscous parameters corresponds to
$z_T=0.49^{+0.064}_{-0.066}.$}
\end{figure}
The variation of $q$ with $z$ for the two sets of limiting
conditions of the viscous parameters are shown in figures
\ref{fig:deceleration1} and \ref{fig:deceleration2}. The evolution
corresponding to the best estimates from both limiting conditions
are identical as it is clear from the figures. When all the bulk
viscous parameters are zero, the deceleration parameter $q=1/2$,
which corresponds to a decelerating matter dominated universe with
null bulk viscosity.

The present value of the deceleration parameter corresponds to $z=0$
or $a=1$ is,
\begin{equation}
\label{q1}
q_0=q(a=1)=\frac{1-(\tilde{\zeta}_{0}+\tilde{\zeta}_{1})}{2-\tilde{\zeta}_{2}}.
\end{equation}
This equation shows that for
$\tilde{\zeta}_{0}+\tilde{\zeta}_{1}=1$, the deceleration parameter
$q=0$. This implies that the transition into the accelerating phase
would occur at the present time and is true for both the cases of
the parameters.

For the first case of limiting conditions of the parameters
(\ref{condition1}) with $\tilde{\zeta}_{0}>0$ and
$\tilde{\zeta}_{2}<2$, the current deceleration parameter $q_0<0$ if
$\tilde{\zeta}_{0}+\tilde{\zeta}_{1}>1,$ implying that the present
universe is in the accelerating epoch and it entered this epoch at
an early stage. But $q_0>0$ if
$\tilde{\zeta}_{0}+\tilde{\zeta}_{1}<1,$ implying that the present
universe is decelerating and it will be entering the accelerating
phase at a future time, see figure \ref{fig:deceleration1} which
shows the behavior of $q$ with $z.$ For the best estimate of the
bulk viscous parameters, the behavior of $q$ (figure
\ref{fig:deceleration1}) shows that the universe transit from
decelerated to accelerated epoch at a recent past. The best estimate
of the bulk viscous parameters corresponding to the first limiting
case, equation \ref{condition1} were extracted using the Supernova
data and are
$(\tilde{\zeta}_{0}=7.83,\tilde{\zeta}_{1}=-5.13,\tilde{\zeta}_{2}=-0.51)$
(see Table \ref{tab:1}), which indicate that
$\tilde{\zeta}_{0}+\tilde{\zeta}_{1}>1.$ So the model predicts a
universe which is accelerating at present and has entered this phase
of accelerating expansion at a recent past.

For the second case of limiting conditions of the viscous parameters
(\ref{condition2}) with $\tilde{\zeta}_{0}<0$ and
$\tilde{\zeta}_{2}>2$, the current deceleration parameter
 $q_0>0$ if $\tilde{\zeta}_{0}+\tilde{\zeta}_{1}>1,$ implies that the present universe is in the
 decelerating epoch and it will be entering the accelerating phase
at a future time, see figure \ref{fig:deceleration2} which shows the
behavior of $q$ with $z.$ But $q_0<0$ if
$\tilde{\zeta}_{0}+\tilde{\zeta}_{1}<1,$ implying that the present
universe is accelerating and it entered this phase at an early time.
From the behavior of $q$ (figure \ref{fig:deceleration2}) for the
best estimate of the bulk viscous parameters corresponding to the
second limiting condition, equation \ref{condition2}, it is clear
that the transition of the universe from the decelerated to
accelerated epoch was in the recent past. The best estimate of the
bulk viscous parameters in this case are
$(\tilde{\zeta}_{0}=-4.68,\tilde{\zeta}_{1}=4.67,\tilde{\zeta}_{2}=3.49)$
(see Table \ref{tab:1}), which indicate that
$\tilde{\zeta}_{0}+\tilde{\zeta}_{1}<1.$ So, for this case also, the
model predicts a universe which is accelerating at present and has
entered this phase of accelerating expansion at a recent past.

These results confirm the earlier conclusion with respect to the
behavior of $d^{2}a/dy^{2}$. For the best estimated values of the
bulk viscous parameters, the present value of the deceleration
parameter is found to be about $-0.68\pm0.06$ and
$-0.68^{+0.066}_{-0.05}$ corresponding to the first and second
limiting conditions respectively (see equation (\ref{q1})). This is
comparable with the observational results on the present value of q,
which is around $-0.64\pm0.03$ \cite{tegmark2,wmap1}. The transition
red shift, at which $q$ enters the negative value region,
corresponding to an accelerating epoch, is found to be
$z_{T}=0.49^{+0.075}_{-0.057}$ for the first case of limiting
conditions of the bulk viscous parameters and
$z_{T}=0.49^{+0.064}_{-0.066}$ for the second case of limiting
conditions of the bulk viscous parameters (see equation
(\ref{transz}) and figures \ref{fig:deceleration1} and
\ref{fig:deceleration2}). An analysis of the $\Lambda$CDM model with
combined SNe+CMB data gives the transition red shift range as
$z_{T}=0.45 - 0.73$ \cite{alam1}. So the transition red shift
predicted by the present model is agreeing only with the lower limit
of the corresponding $\Lambda$CDM range, and hence can be hardly
considered as a good agreement.
\subsection{Equation of state} \label{sec:5} An accelerated expansion
of the universe is possible only if the effective equation of state
parameter,
 $\omega<-1/3,$ or equivalently, $3\omega+1<0$.
The equation of state can be obtained using \cite{praseetha1},
\begin{equation}
\omega=-1-\frac{1}{3}\frac{d \ln h^{2}}{dx}
\end{equation}
where $x=\ln a$ and $h=\frac{H}{H_{0}}$. Using equation
(\ref{hubbleina}) we get the equation of state as,
\begin{equation}
\omega=-1-\frac{2}{3(2-\tilde{\zeta}_{2})}\left[\tilde{\zeta}_{1}+\tilde{\zeta}_{2}-3+\frac{\tilde{\zeta}_{0}}{h}\right]
\end{equation}
The present value of the equation of state parameter $\omega_0$, can
be obtained by taking $h=1$. The condition for acceleration of the
present universe can then be represented as,
\begin{equation}
3\omega_0+1=-2\left(\frac{\tilde{\zeta}_{0}+\tilde{\zeta}_{1}-1}{2-\tilde{\zeta}_{2}}\right)<0
\end{equation}
For the first case of parameters with $\tilde{\zeta}_{0}>0$,
$\tilde{\zeta}_{2}<2$, this condition is satisfied if
$\tilde{\zeta}_{0}+\tilde{\zeta}_{1}>1$ and for the second case with
$\tilde{\zeta}_{0}<0$, $\tilde{\zeta}_{2}>2$, this will be satisfied
if $\tilde{\zeta}_{0}+\tilde{\zeta}_{1}<1$. These conditions are
compatible with that arrived in the analysis of deceleration
parameter in section \ref{sec:4}.
\begin{figure}
\centering
\includegraphics[scale=0.6]{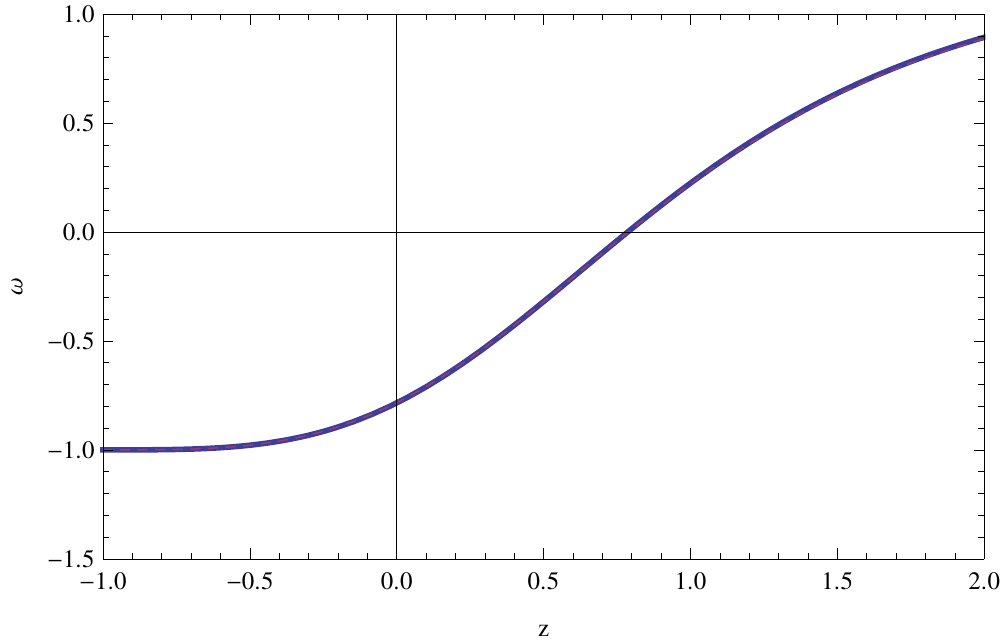}
\caption{\label{fig:equation} Evolution of the equation of state
parameter with red shift for the best estimates of the bulk viscous
parameters. It is found that the evolution of $\omega$ are identical
for the best estimates from both the limiting conditions.}
\end{figure}

The evolution of the equation of state parameter with red shift for
both the sets of the best fit values of the bulk viscous parameters
are found to be identical and is shown in figure \ref{fig:equation}.
It is clear from the figure that as $z\rightarrow-1$
($a\rightarrow\infty$), $\omega\rightarrow-1$ in the future which
corresponds to the de Sitter universe and also coincides with that
of the future behavior of the $\Lambda$CDM model \cite{alcaniz1},
and also resembles the behavior of some scalar field models
\cite{cope}. Since it is not crossing the phantom divide
$\omega\leq-1$, the model is free from big rip singularity or little
rip \cite{Brevik2}. The present value of the equation of state
parameter is around $\omega_{0}\sim-0.78^{+0.03}_{-0.045}$ and
$\omega_{0}\sim-0.78^{+0.037}_{-0.043}$ for the best estimate of
viscosity parameters corresponding to the first and second limiting
conditions, respectively. This value is comparatively higher than
that predicted by the joint analysis of WMAP+BAO+$H_{0}$+SN data,
which is around $-0.93$ \cite{komatsu1,chimanto1}.

\subsection{Evolution of matter density}
\label{sec:6} From the  Friedmann equation (\ref{friedmann}) and the
Hubble parameter (\ref{hubbleina}) we obtain the mass density
parameter $\Omega_{m}$ as,
\begin{equation}
\Omega_{m}(a)=\left[a^{\frac{\tilde{\zeta}_{12}-3}{2-\tilde{\zeta}_{2}}}\left[1+\frac{\tilde{\zeta}_{0}}{\tilde{\zeta}_{12}-3}\right]-\frac{\tilde{\zeta}_{0}}{\tilde{\zeta}_{12}-3}\right]^{2}
\end{equation}
\begin{figure}
\centering
\includegraphics[scale=0.6]{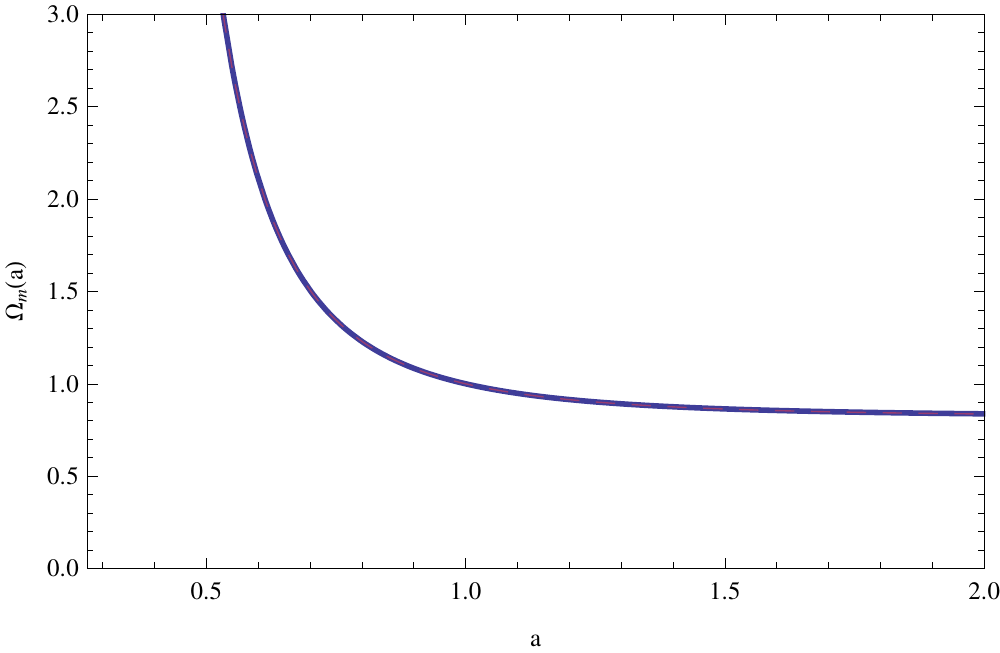}
\caption{\label{fig:matter} Evolution of the mass density parameter
with scale factor for the best estimated values of the bulk viscous
parameters. It is found that the variation of the mass density
coincides for the best estimated values from the two limiting
conditions.}
\end{figure}
where, $\Omega_{m}=\frac{\rho_{m}}{\rho_{crit}}$ and
$\rho_{crit}=3H_{0}^{2}$ is the critical density. If
$\tilde{\zeta}_{0}=\tilde{\zeta}_{1}=\tilde{\zeta}_{2}=0$, the mass
density parameter reduces to $\Omega_{m}\sim a^{-3}$, which
corresponds to the matter dominated universe with null bulk
viscosity. The evolution of the mass density parameter for the best
estimated values corresponding to the two limiting conditions are
shown in figure \ref{fig:matter} and it is clear that their
evolutions are coinciding with each other. As $a\rightarrow 0$, the
matter density diverges. Figure \ref{fig:matter} also indicating the
same, which is a clear indication of the existence of the Big-Bang
at the origin of the universe.

\subsection{The curvature scalar} \label{sec:7} The curvature scalar
is the parameter used to confirm the presence of singularities in
the model. For a flat universe, the curvature scalar is defined as,
\begin{equation}
R=6\left[\frac{\ddot{a}}{a}+H^{2}\right].
\end{equation}
Using equations (\ref{derihubble}), (\ref{dimen}), (\ref{hubbleina})
and (\ref{adouble}), we obtain the curvature scalar as,
\begin{equation}
\begin{aligned}
\begin{split}
R(a)=&\frac{6H_{0}^{2}}{(2-\tilde{\zeta}_{2})(\tilde{\zeta}_{12}-3)^{2}}[2\tilde{\zeta}_{0}^{2}(2-\tilde{\zeta}_{2})
+(\tilde{\zeta}_{0}+\tilde{\zeta}_{12}-3)
\\&a^{\frac{\tilde{\zeta}_{12}-3}{2-\tilde{\zeta}_{2}}}
[(\tilde{\zeta}_{1}-\tilde{\zeta}_{2}+1)(\tilde{\zeta}_{0}+\tilde{\zeta}_{12}-3)
a^{\frac{\tilde{\zeta}_{12}-3}{2-\tilde{\zeta}_{2}}}-\\&
\tilde{\zeta}_{0}(\tilde{\zeta}_{1}-3\tilde{\zeta}_{2}+5)]].
\end{split}
\end{aligned}
\end{equation}
\begin{figure}
\centering
\includegraphics[scale=0.6]{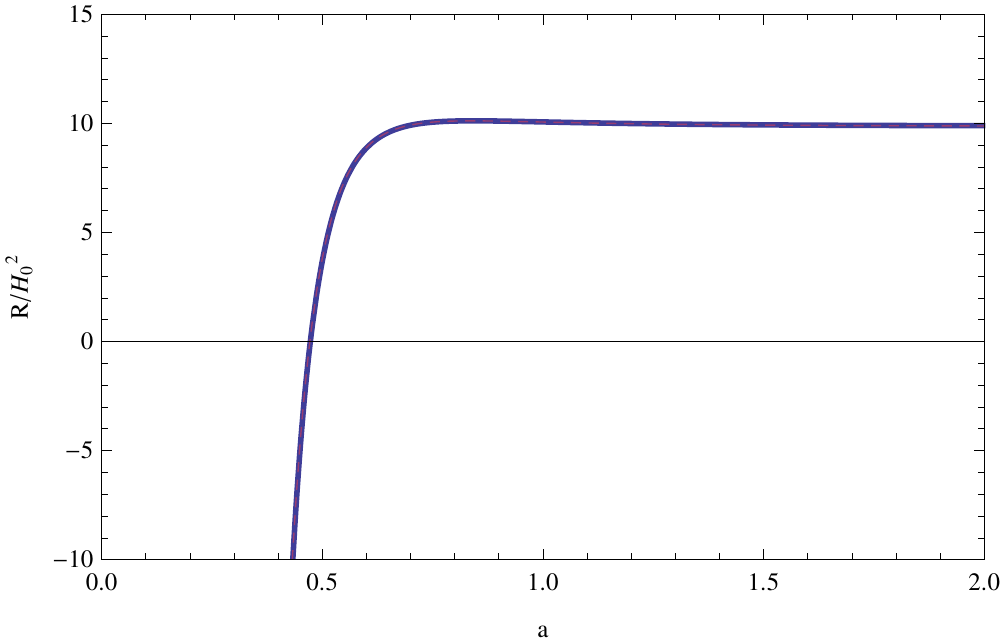}
\caption{\label{fig:curvature1} Evolution of the curvature scalar
with scale factor for the best estimate parameters. It is found that
the evolution of the curvature scalar are identical for the best
estimated values from the two limiting conditions.}
\end{figure}
From the above equation it is clear that as $a\rightarrow0$,
$R\rightarrow\infty.$ The evolution of the curvature scalar for both
the cases of best fit of the parameters coincides with each other as
shown in figure \ref{fig:curvature1}. The behavior of $R$ shows that
the curvature scalar diverges as $a\to0.$ This indicates the
existence of Big-Bang at the origin of the universe.

\section{Entropy and second law of thermodynamics}
\label{sec:8} 
\begin{figure}
\centering
\includegraphics[scale=0.6]{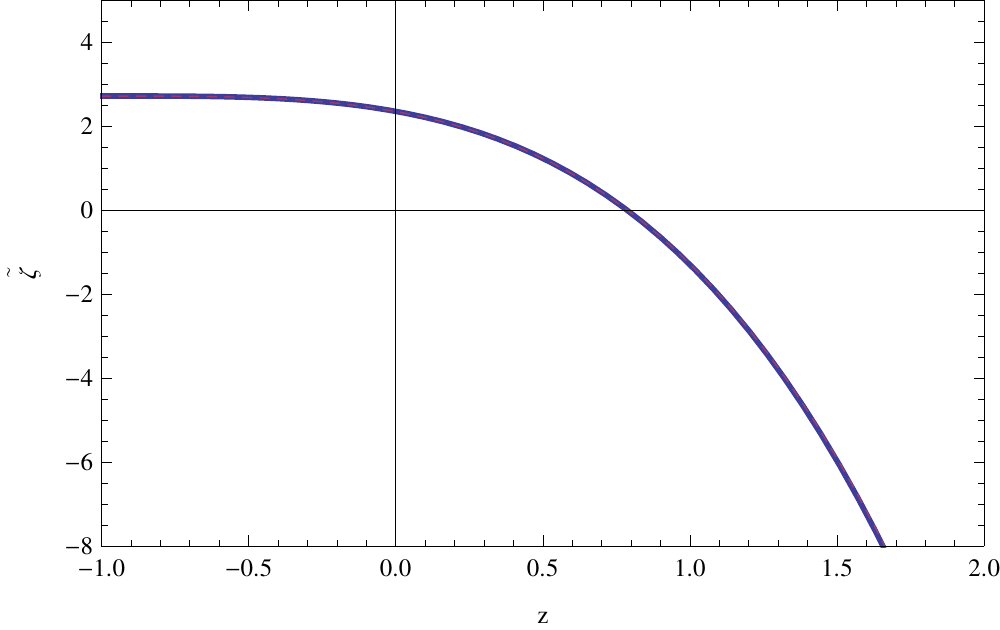}
\caption{\label{fig:entropy1} Evolution of the total dimensionless
bulk viscous parameter with respect to the red shift for the best
estimated values corresponding to the two limiting conditions.
$\tilde{\zeta}$ is positive for $z\leq0.8.$}
\end{figure}

In the FLRW space-time, the law of generation of the
local entropy is given as \cite{weinberg2}
\begin{equation}
\label{entropy}
T\nabla_{\nu}s^{\nu}=\zeta(\nabla_{\nu}u^{\nu})^{2}=9H^{2}\zeta
\end{equation}
where $T$ is the temperature and $\nabla_{\nu}s^{\nu}$ is the rate
of generation of entropy in unit volume. The second law of
thermodynamics will be satisfied if,
\begin{equation}
T\nabla_{\nu}s^{\nu}\geq 0
\end{equation}
which implies from equation (\ref{entropy}) that
\begin{equation}
\label{en} \zeta\geq 0.
\end{equation}
Using equations (\ref{derihubble}) and (\ref{hubbleina}), the total
dimensionless bulk viscous parameter (equation (\ref{zeta})), can be
obtained as

\begin{equation}
\label{zeta1}
\tilde{\zeta}(a)=\frac{1}{2-\tilde{\zeta}_{2}}\left[2\tilde{\zeta}_{0}+\left(2\tilde{\zeta}_{1}-\tilde{\zeta}_{2}\right)
\frac{H}{H_{0}}\right],
\end{equation}
where  $\tilde{\zeta}=\frac{3\zeta}{H_{0}}$, the total dimensionless
bulk viscous parameter. We have studied the evolution of
$\tilde{\zeta}$ using the best estimated values for both cases of
parameters and found that the evolution of the total bulk viscous
parameter are coinciding for both the cases as shown in figure
\ref{fig:entropy1}. The figure also shows that the total bulk
viscous coefficient is evolving continuously from the negative value
region to a positive region. When $z\leq0.8$, the total bulk viscous
parameter becomes positive. This means that the rate of entropy
production is negative in the early epoch and positive in the later
epoch. Hence the local second law is violated in the early epoch and
is obeyed in the later epoch. This seems to be a drawback of the
present model. However, it can be considered as a theoretical
possibility \cite{Brevik}. In an absolute way the status of the
second law of thermodynamics should be considered along with the
accounting of the entropy generation from the horizon. In that
circumstances, the second law becomes the generalized second law of
thermodynamics, which state that the total entropy of the fluid
components of the universe plus that of the horizon should never
decrease \cite{gibbons1,tkm1}. In the present model this means the
rate of entropy change of the bulk viscous matter and that of the
horizon must be greater than zero.
\begin{equation}
\frac{d}{dt}\left(S_{m}+S_{h}\right)\geq0
\end{equation}
where, $S_{m}$ is the entropy of the matter and $S_{h}$ is that of
the horizon. For a flat FLRW universe, the apparent horizon radius
is given as \cite{Sheykhi1}
\begin{equation}
\label{apparentradius} r_{A}=\frac{1}{H}.
\end{equation}
The entropy associated to the apparent horizon is \cite{davis1},
\begin{equation}
\label{horizonentropy} S_{h}=2\pi A=8\pi^{2}r_{A}^{2}
\end{equation}
where $A=4\pi r_{A}^{2}$ is the area of the apparent horizon and we
have assumed $8\pi G=1$. Using the first Friedmann equation and
equations (\ref{p}), (\ref{conser}), (\ref{z})
 and (\ref{apparentradius}), we obtain,
\begin{equation}
\label{radius}
\dot{r}_{A}=\frac{1}{2}r_{A}^{3}H\left[-H(\tilde{\zeta}_{0}H_{0}
+\tilde{\zeta}_{1}H+\tilde{\zeta}_{2}(\frac{\dot{H}}{H}+H))+\rho_{m}\right]
\end{equation}
The temperature of the apparent horizon can be defined as
\cite{setare1}
\begin{equation}
\label{temperature} T_{h}=\frac{1}{2\pi
r_{A}}\left(1-\frac{\dot{r}_{A}}{2Hr_{A}}\right).
\end{equation}
Using equations (\ref{horizonentropy}), (\ref{radius}) and
(\ref{temperature}), we arrive
\begin{equation}
\begin{split}
\label{1} T_{h}\dot{S}_{h}=4\pi
r_{A}^{3}H\left[\rho_{m}-H(\tilde{\zeta}_{0}H_{0}+\tilde{\zeta}_{1}H+\tilde{\zeta}_{2}
(\frac{\dot{H}}{H}+H))\right]\\
\left[1-\frac{\dot{r}_{A}}{2Hr_{A}}\right].
\end{split}
\end{equation}
The change in entropy of the viscous matter inside the apparent
horizon can be obtained using the Gibbs equation,
\begin{equation}
\label{gibbs} T_{m}dS_{m}=d(\rho_{m}V)+P^{*}dV
\end{equation}
where $T_{m}$ is the temperature of the bulk viscous matter,
$V=\frac{4}{3}\pi r_{A}^{3}$ is the volume enclosed by the apparent
horizon. Using equations (\ref{p}) and (\ref{z}), the Gibbs equation
becomes
\begin{equation}
\label{gibbs1}
T_{m}dS_{m}=Vd\rho_{m}+(\rho_{m}-H(\tilde{\zeta}_{0}H_{0}+\tilde{\zeta}_{1}H+\tilde{\zeta}_{2}(\frac{\dot{H}}{H}+H)))dV.
\end{equation}
Under equilibrium conditions, the temperature $T_{m}$ of the viscous
matter and that of the horizon $T_{h}$ are equal, $T_{m}=T_{h}$.
Then the Gibbs equation (\ref{gibbs1}) becomes
\begin{equation}
\begin{split}
\label{2} T_{h}\dot{S}_{m}=4\pi
r_{A}^{3}H\left[H(\tilde{\zeta}_{0}H_{0}+\tilde{\zeta}_{1}H+\tilde{\zeta}_{2}
(\frac{\dot{H}}{H}+H)-\rho_{m})\right] \\ +4\pi
r_{A}^{2}\dot{r}_{A}\left[\rho_{m}-H(\tilde{\zeta}_{0}H_{0}+\tilde{\zeta}_{1}H+\tilde{\zeta}_{2}
(\frac{\dot{H}}{H}+H))\right].
\end{split}
\end{equation}
Adding equations (\ref{1}) and (\ref{2}), we get
\begin{equation}
T_{h}(\dot{S}_{h}+\dot{S}_{m})=\frac{A}{4} Hr_{A}^{3}
[\rho_{m}-H(\tilde{\zeta}_{0}H_{0}+\tilde{\zeta}_{1}H+\tilde{\zeta}_{2}
({\dot{H}\over H}+H))]^{2}.
\end{equation}
$A$, the area of the apparent horizon, $H$, the Hubble parameter and
the radius $r_{A}$ are always positive, therefore,
$\dot{S}_{h}+\dot{S}_{m}\geq0$ for a given temperature. This means
that the generalized second law (GSL) is always valid. Hence the
decrease in the entropy of the viscous matter is compensated by the
increase in the entropy of the horizon. Even though the violation of
the local second law of thermodynamics can be considered as a draw
back of this model, the validity of the Generalized second law for
the entire causal region of the universe may safe guard the model.

\section{Statefinder analysis}
\label{sec:9} 
\begin{figure}
\centering
\includegraphics[scale=0.6]{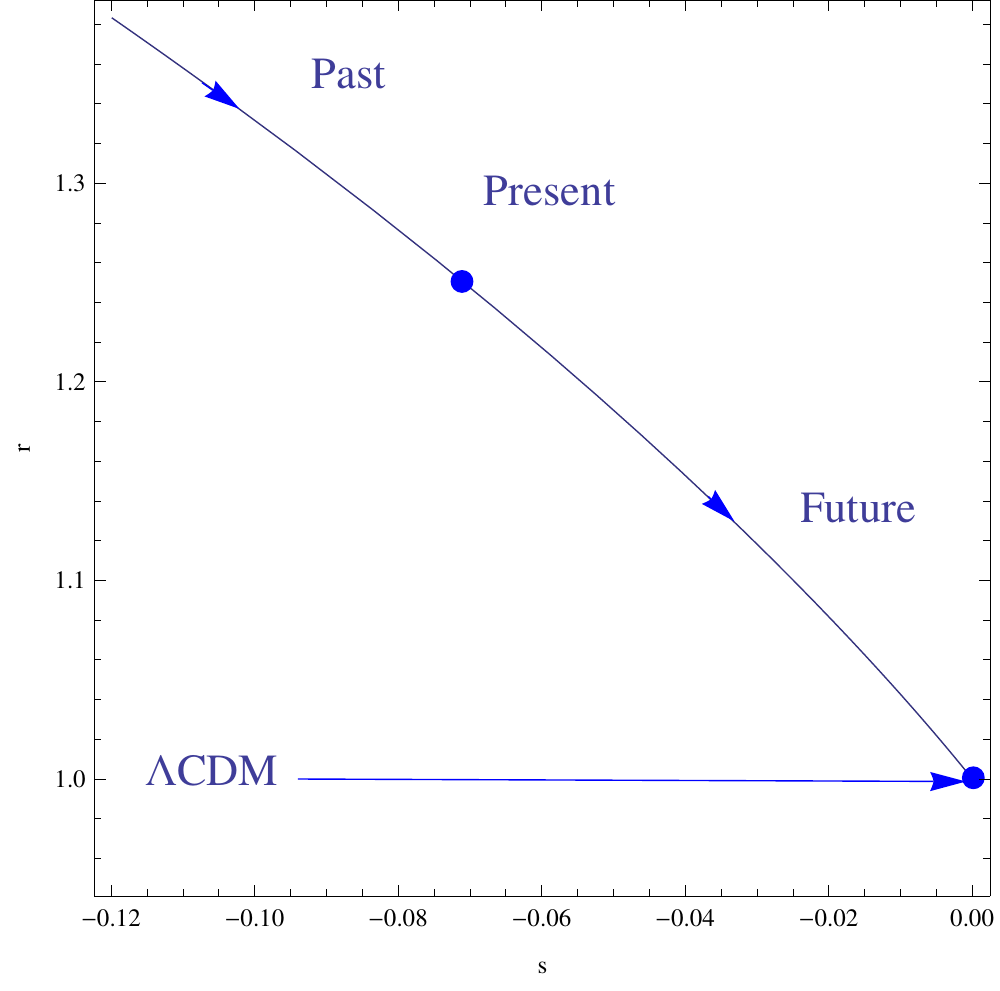}
\caption{\label{fig:statefinder1} The evolution of the model in the
r-s plane for the best estimates of the parameters. The curves are
coinciding with each other for the best estimated values of the
parameters from both the limiting conditions.}
\end{figure}

In this section, we present our analysis on comparing
the present model with other standard models of dark energy. We have
used the statefinder parameter diagnostic introduced by Sahni et al
\cite{sahni1}. The statefinder is a geometrical diagnostic tool
which allows us to characterize the properties of dark energy in a
model-independent manner. The statefinder parameters $\{r,s\}$ are
defined as,
\begin{equation}
r=\frac{\dddot{a}}{aH^3}, \ \ \ \ \ \
s=\frac{r-1}{3\left(q-\frac{1}{2}\right)}.
\end{equation}
In terms of $h=\frac{H}{H_0}$, $r$ and $s$ can be written as
\begin{equation}
r=\frac{1}{2h^{2}}\frac{d^{2}h^2}{dx^2}+\frac{3}{2h^2}\frac{dh^2}{dx}+1
\end{equation}
\begin{equation}
s=-\frac{\frac{1}{2h^{2}}\frac{d^{2}h^2}{dx^2}+\frac{3}{2h^2}\frac{dh^2}{dx}}{\frac{3}{2h^2}\frac{dh^2}{dx}+\frac{9}{2}}.
\end{equation}
Using the expression for $h$ from equation (\ref{hubbleina}), these
parameters become,
\begin{equation}
\begin{split}
r=\frac{(\tilde{\zeta}_0+\tilde{\zeta}_{12}-3)(\tilde{\zeta}_{12}-3)}{h^2(2-\tilde{\zeta}_2)^2}
a^{\frac{\tilde{\zeta}_{12}-3}{2-\tilde{\zeta}_2}}[2h+\frac{\tilde{\zeta}_0}{\tilde{\zeta}_{12}-3}]+\\
\frac{3(\tilde{\zeta}_0+\tilde{\zeta}_{12}-3)}{h(2-\tilde{\zeta}_2)}a^{\frac{\tilde{\zeta}_{12}-3}{2-\tilde{\zeta}_2}}+1
\end{split}
\end{equation}
\begin{equation}
s=\frac{\frac{(\tilde{\zeta}_0+\tilde{\zeta}_{12}-3)(\tilde{\zeta}_{12}-3)}{h^2(2-\tilde{\zeta}_2)^2}
a^{\frac{\tilde{\zeta}_{12}-3}{2-\tilde{\zeta}_2}}[2h+\frac{\tilde{\zeta}_0}{\tilde{\zeta}_{12}-3}]+
\frac{3(\tilde{\zeta}_0+\tilde{\zeta}_{12}-3)}{h(2-\tilde{\zeta}_2)}a^{\frac{\tilde{\zeta}_{12}-3}{2-\tilde{\zeta}_2}}}{
\frac{3(\tilde{\zeta}_0+\tilde{\zeta}_{12}-3)}{h(2-\tilde{\zeta}_2)}a^{\frac{\tilde{\zeta}_{12}-3}{2-\tilde{\zeta}_2}}+\frac{9}{2}}.
\end{equation}
The above equations show that in the limit $a\to \infty$, the
statefinder parameters $\{r,s\}\to \{1,0\}$, a value corresponding
to the $\Lambda$CDM model of the universe. So the present model
resembles the $\Lambda$CDM model in the future. The $\{r,s\}$ plane
trajectory of the model is shown in figure \ref{fig:statefinder1}.
The trajectories are coinciding with each other for the best
estimates from both the sets of the limiting conditions of the
parameters. The trajectory in the $\{r,s\}$ plane are lying in the
region $r>1,s<0$, a feature similar to the generalized Chaplygin gas
model of dark energy \cite{wu1}. The present model can also be
discriminated from the Holographic dark energy model with event
horizon as the I.R. cut off, in which the $r-s$ evolution starts
from a region $r\sim1,s\sim2/3$ and end on the $\Lambda$CDM point
\cite{liu}. The present position of the universe dominated by the
bulk viscous matter is noted in the plot and it corresponds to
$\{r_0,s_0\}=\{1.25,-0.07\}$. This means that the present model is
distinguishably different from the $\Lambda$CDM model.

\section{Parameter estimation using type Ia Supernovae data}
\label{sec:estmation}
 In this section we have obtained best
fit of the parameters, $\tilde{\zeta}_{0}$, $\tilde{\zeta}_{1}$,
$\tilde{\zeta}_{2}$ and $H_0$ using the type Ia Supernovae
observations. The goodness-of-fit of the model is obtained by the
$\chi^{2}$-minimization. We did the statistical analysis using the
Supernova Cosmology Project (SCP) ``Union" SNe Ia data set
\cite{kowalski1}, composed of 307 type Ia Supernovae from 13
independent data sets.

In a flat universe, the luminosity distance $d_{L}$ is defined as
\begin{equation}
d_{L}(z,\tilde{\zeta}_{0},\tilde{\zeta}_{1},\tilde{\zeta}_{2},H_{0})=c(1+z)\int_{0}^{z}\frac{dz'}{H(z',\tilde{\zeta}_{0},\tilde{\zeta}_{1},\tilde{\zeta}_{2},H_{0})}
\end{equation}
where
$H(z,\tilde{\zeta}_{0},\tilde{\zeta}_{1},\tilde{\zeta}_{2},H_{0})$
is the Hubble parameter and $c$ is the speed of light. The
theoretical distance moduli $\mu_{t}$ for the k-th Supernova with
redshift $z_{k}$ is given as,
\begin{equation}
\begin{aligned}
\begin{split}
\mu_{t}(z_{k},\tilde{\zeta}_{0},\tilde{\zeta}_{1},\tilde{\zeta}_{2},H_{0})&=m-M
\\&=5\log_{10}[\frac{d_{L}(z_{k},\tilde{\zeta}_{0},\tilde{\zeta}_{1},\tilde{\zeta}_{2},H_{0})}{Mpc}]+25
\end{split}
\end{aligned}
\end{equation}
where, $m$ and $M$ are the apparent and absolute magnitudes of the
SNe respectively. Then we can construct $\chi^{2}$ function as,
\begin{equation}
\chi^{2}(\tilde{\zeta}_{0},\tilde{\zeta}_{1},\tilde{\zeta}_{2},H_{0})\equiv
\sum^{n}_{k=1}\frac{\left[\mu_{t}(z_{k},\tilde{\zeta}_{0},\tilde{\zeta}_{1},\tilde{\zeta}_{2},H_{o})-\mu_{k}\right]^{2}}{\sigma_{k}^{2}}
\end{equation}
where $\mu_{k}$ is the observational distance moduli for the k-th
Supernova, $\sigma_{k}^{2}$ is the variance of the measurement and
$n$ is the total number of data, here $n=307$. The $\chi^{2}$
function, thus obtained is then minimized to obtain the best
estimate of the parameters, $\tilde{\zeta}_{0}$,
$\tilde{\zeta}_{1}$, $\tilde{\zeta}_{2}$ and $H_0$. From the
behavior of scale factor and other cosmological parameters, we found
that there exists two possible sets of conditions which describes a
universe having a Big-Bang at the origin, then entering an early
stage of decelerated expansion followed by acceleration. These two
sets of conditions are mentioned in section \ref{sec:3}. We have
used these two conditions separately in minimizing the $\chi^{2}$
function. This leads to two sets of values for the best estimates of
the parameters $\tilde{\zeta}_{0}$, $\tilde{\zeta}_{1}$,
$\tilde{\zeta}_{2}$ but $H_0$ is same in both the cases. In addition
to $H_0$, the other cosmological parameters, scale factor,
deceleration parameter, equation of state parameter, matter density
and curvature scalar are all showing identical behavior for both the
sets of best fit of parameters. The values of the parameters are
given in Table \ref{tab:1}. Inorder to compare the results of the
present model, we have also estimated the values for $\Lambda$CDM
model using the same data set and the results are also shown in
Table \ref{tab:1}. We find that the values of $H_{0}$ and
Goodness-of-fit $\chi^{2}_{d.o.f}$ for $\Lambda$CDM model are very
close to those obtained from the present bulk viscous model. The
value of the present Hubble parameter, $H_0$ for both the cases of
parameters are found to be $70.49$ $km s^{-1}Mpc^{-1}$, which is in
close agreement with the corresponding WMAP value
($H_{0}=70.5\pm1.3$ $km s^{-1}Mpc^{-1}$) \cite{wmap1}.

\begin{table}
\caption{Best estimates of the Bulk viscous parameters and $H_{0}$
and also $\chi^{2}$ minimum value for the two cases of the bulk
viscous matter dominated universe.
$\chi^{2}_{d.o.f}=\frac{\chi^{2}_{min}}{n-m}$, where $n=307$, the
number of data and $m=3$, the number of parameters in the model. For
the best estimation we have use SCP ``Union" 307 SNe Ia data sets.
We have also shown the best estimates for the $\Lambda$CDM model for
comparison, where $\Omega_{m0}$ is the present mass density
parameter. The subscript d.o.f stands for degrees of freedom.}
\begin{tabular}{|m{1.5cm}|p{2cm}|p{2cm}|c|} \hline
Model $\rightarrow$ & Bulk viscous model with
$\tilde{\zeta}_{0}>0,\tilde{\zeta}_{0}+\tilde{\zeta}_{12}<3,\tilde{\zeta}_{12}<3,\tilde{\zeta}_{2}<2$
& Bulk viscous model with
$\tilde{\zeta}_{0}<0,\tilde{\zeta}_{0}+\tilde{\zeta}_{12}>3,\tilde{\zeta}_{12}>3,\tilde{\zeta}_{2}>2$ & $\Lambda$CDM \\
\hline $\tilde{\zeta}_{0}$ & 7.83 & -4.68 & - \\ \hline
$\tilde{\zeta}_{1}$ & $-5.13^{+0.056}_{-0.060}$ &
$4.67^{+0.04}_{-0.03}$ & - \\ \hline $\tilde{\zeta}_{2}$ &
$-0.51^{+0.13}_{-0.14}$ &
$3.49^{+0.089}_{-0.071}$ & - \\ \hline $\Omega_{m0}$ & 1 & 1 & 0.316 \\
\hline $H_{0}$ & 70.49 & 70.49 &
70.03 \\ \hline $\chi^{2}_{min}$ & 310.54 & 310.54 & 311.93 \\
\hline $\chi^{2}_{d.o.f}$ & 1.02 & 1.02 & 1.02 \\ \hline
\label{tab:1}
\end{tabular}
\end{table}

\begin{figure}
\centering
\includegraphics[scale=0.6]{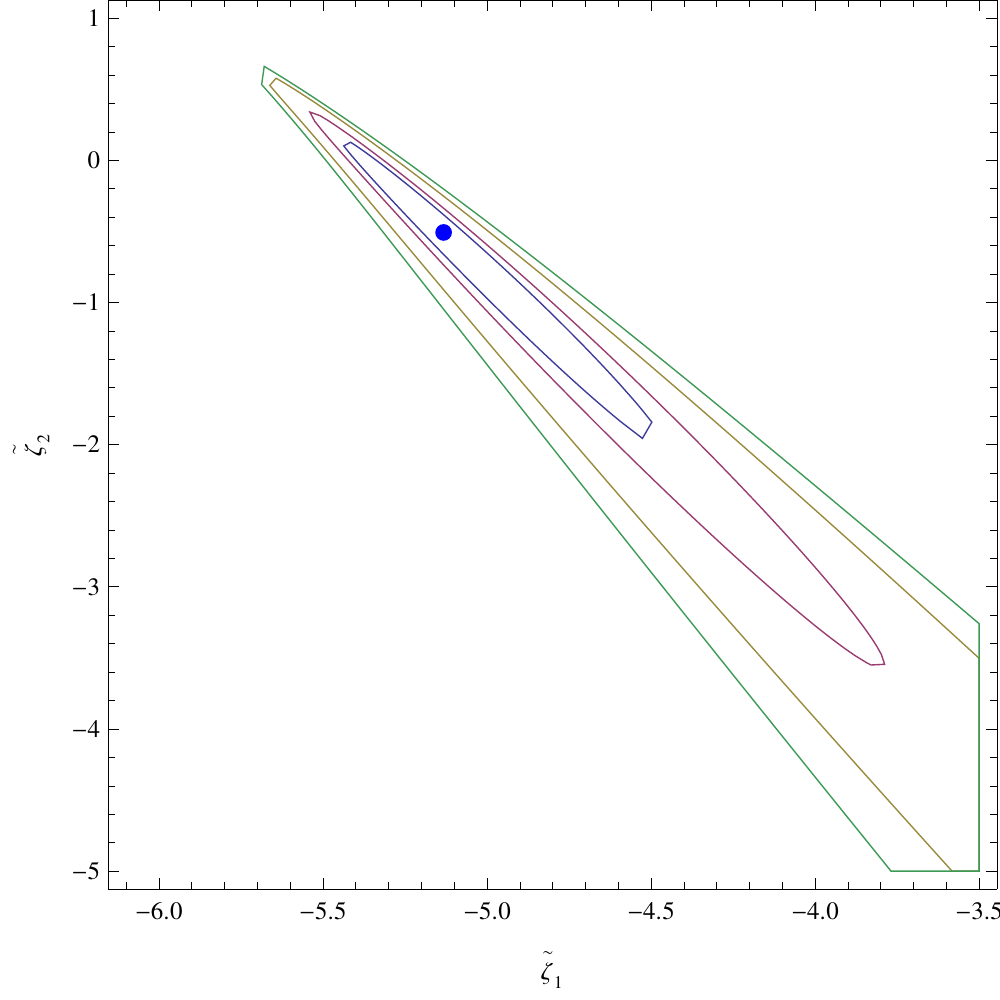}
\caption{\label{fig:confidence1} Confidence intervals for the
parameters $(\tilde{\zeta}_{1},\tilde{\zeta}_{2})$, for the first
set of limiting conditions, for the bulk viscous matter dominated
universe using the SCP ``Union" data set composed of 307 data
points. The best estimated values of the parameters are
$\tilde{\zeta}_{1}=-5.13^{+0.056}_{-0.06}$ and $\tilde{\zeta}_{2}=
-0.51^{+0.13}_{-0.14}$ and are indicated by the point. The
confidence intervals shown corresponds to 68.3\%, 95.4\%, 99.73\%
and 99.99\% of probabilities. }
\end{figure}
\begin{figure}
\centering
\includegraphics[scale=0.6]{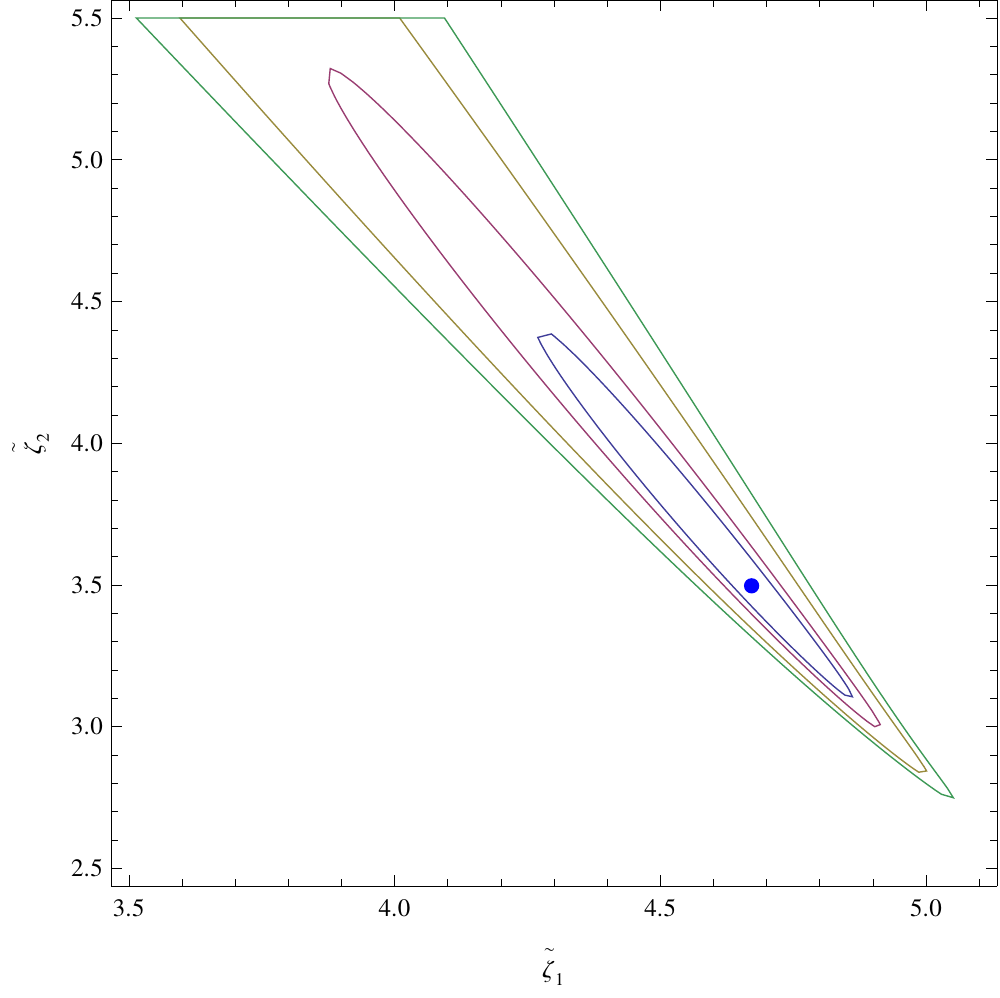}
\caption{\label{fig:confidence2} Confidence intervals for the
parameters $(\tilde{\zeta}_{1},\tilde{\zeta}_{2})$, for the second
set of limiting conditions, for the bulk viscous matter dominated
universe using the SCP ``Union" data set composed of 307 data
points. The best estimated values of the parameters are
$4.67^{+0.04}_{-0.03}$ and $3.49^{+0.089}_{-0.071}$ and are
indicated by the point. The confidence intervals shown corresponds
to 68.3\%, 95.4\%, 99.73\% and 99.99\% of probabilities.}
\end{figure}

We have constructed the confidence interval plane for the bulk
viscous parameters $(\tilde{\zeta}_{1},\tilde{\zeta}_{2})$ by
keeping $\tilde{\zeta}_{0}$ as a constant equal to its best
estimated value obtained by minimizing the $\chi^2$ function. From
figure \ref{fig:confidence1}, corresponding to the first set of
limiting conditions, and figure \ref{fig:confidence2}, corresponding
to the second set of limiting conditions, it is seen that the
fitting of the confidence intervals corresponding to $99.73\%$ and
$99.99\%$ probabilities are poor. But the confidence intervals
corresponding to $68.3\%$ and $95.4\%$ probabilities are showing a
fairly good behavior. From the equation of the total bulk viscous
coefficient (equation (\ref{zeta1})) it can be easily verified that
the present value of the total viscosity coefficient is positive in
the region of confidence interval.

For the first case of parameters with $\tilde{\zeta}_{0}>0$, it is
found that $\tilde{\zeta}_{1}=-5.13^{+0.056}_{-0.06}$ and
$\tilde{\zeta}_{2}= -0.51^{+0.13}_{-0.14}$, for
$\tilde{\zeta}_{0}=7.83$ with $68.3\%$ probability. In the second
case with $\tilde{\zeta}_{0}<0$, the values of $\tilde{\zeta}_{1}$
and $\tilde{\zeta}_{2}$ are obtained as $4.67^{+0.04}_{-0.03}$ and
$3.49^{+0.089}_{-0.071}$, respectively, for
$\tilde{\zeta}_{0}=-4.68$ with $68.3\%$ probability.

\section{Conclusions}
\label{sec:conclu} In this paper, we have carried out a study of the
bulk viscous matter dominated universe with bulk viscosity of the
form
$\zeta=\zeta_{0}+\zeta_{1}\frac{\dot{a}}{a}+\zeta_{2}\frac{\ddot{a}}{\dot{a}}.$
This model automatically solves the coincidence problem because the
bulk viscous matter simultaneously represents dark matter and dark
energy and causes recent acceleration. We have identified two
possible limiting conditions for bulk viscous parameters where the
universe begins with a Big-Bang, followed by decelerated expansion
in the early times and then making a transition to the accelerated
epoch at recent past. These conditions corresponds to
$(\tilde{\zeta}_{0}>0, \tilde{\zeta}_{0}+\tilde{\zeta}_{12}<3,
\tilde{\zeta}_{12}<3, \tilde{\zeta}_{2}<2)$ and
$(\tilde{\zeta}_{0}<0, \tilde{\zeta}_{0}+\tilde{\zeta}_{12}>3,
\tilde{\zeta}_{12}>3$, $\tilde{\zeta}_{2}>2).$

In constraining the parameter we have used SCP ``Union" type Ia
Supernova data set. We have computed the minimum values of $\chi^2$
function by degrees of freedom ($\chi^2_{d.o.f}$) for both cases of
limiting conditions of the bulk viscous parameters and are found to
be very near to one, indicating a reasonable goodness-of-fit. We
have evaluated the best fit values of the three parameters,
$(\tilde{\zeta}_{0}, \tilde{\zeta}_{1},\tilde{\zeta}_{2})$
simultaneously for both cases of limiting conditions of the
parameters and are shown in Table \ref{tab:1}.

For both cases of the best estimate of the bulk viscous parameters,
the evolution of the cosmological parameters: the scale factor,
deceleration parameter, the equation of state parameter, matter
density, curvature scalar are all found to be identical. So these
two sets of best estimated values for the parameters cannot be
distinguished by using the conventional cosmological parameters. By
doing a phase space analysis, it may be possible to distinguish
between these two limiting conditions so as to remove the apparent
degeneracy in the best estimated values of bulk viscous coefficient,
such a work is in progress and will be reported else where.

From the evolution of scale factor, it is found that for the first
limiting conditions of bulk viscous parameters, the transition into
the accelerating epoch would be in the recent past if
$\tilde{\zeta}_{0}+\tilde{\zeta}_{1}>1$. On the other hand if
$\tilde{\zeta}_{0}+\tilde{\zeta}_{1}<1$, the transition takes place
in the future and if, $\tilde{\zeta}_{0}+\tilde{\zeta}_{1}=1$, the
transition takes place at the present time. For the second limiting
conditions of parameters the above conditions are getting reversed
such that when $\tilde{\zeta}_{0}+\tilde{\zeta}_{1}>1$, the
transition will takes place in the future, when
$\tilde{\zeta}_{0}+\tilde{\zeta}_{1}<1$, the transition would occur
in the recent past and when $\tilde{\zeta}_{0}+\tilde{\zeta}_{1}=1$,
the transition takes place at the present time.

We have also estimated the present age of the universe and found to
be around $10.90$ Gyr for the best estimates of the parameters.
Compared to the age predicted from oldest galactic globular clusters
($12.9\pm2.9$ Gyr), the present value is relatively less, but just
within the concordance limit.

The evolution of the deceleration parameter shows that the
transition from the decelerated to the accelerated epoch occurs at
the present time, corresponding to $q=0$ if
$\tilde{\zeta}_{0}+\tilde{\zeta}_{1}=1$, for both sets of limiting
conditions of the parameters. The transition would be in the recent
past, corresponds to $q<0$ at present, if
$\tilde{\zeta}_{0}+\tilde{\zeta}_{1}>1$ for the first set of
limiting conditions and $\tilde{\zeta}_{0}+\tilde{\zeta}_{1}<1$, for
the second set. The transition into the accelerating epoch will be
in the future, corresponds to $q>0$ at present if
$\tilde{\zeta}_{0}+\tilde{\zeta}_{1}<1$ for the first set of
limiting conditions of the parameters and
$\tilde{\zeta}_{0}+\tilde{\zeta}_{1}>1$ for the second set. However,
for the best estimates of viscous parameters from both the limiting
conditions, the behavior of the deceleration parameters are
identical. It is found that for the best estimates, the universe
entered the accelerating phase in the recent past at a red shift
$z_T = 0.49^{+0.075}_{-0.057}$ for the first limiting conditions and
$z_T = 0.49^{+0.064}_{-0.066}$ for the second limiting conditions.
This is found to be agreeing only with the lower limit of the
corresponding $\Lambda$CDM range, $z_T = 0.45-0.73$ \cite{alam1}.
The present value of the deceleration parameter is found to be about
$-0.68^{+0.06}_{-0.06}$ and $-0.68^{+0.066}_{-0.05}$ for the two
cases respectively and is comparable with the observational results
which is around $-0.64\pm0.03.$

We have analyzed the equation of state parameter for the best
estimates of the bulk viscous parameters only. The equation of state
parameter $\omega\to-1$ as $z\to-1,$  which means that the bulk
viscous matter dominated universe behaves like the de Sitter
universe in future. It is also clear that the equation of state
parameter of this model doesn't cross the phantom divide and
thereby, free from big rip singularity. The present value of the
equation of state parameter is around $-0.78^{+0.03}_{-0.045}$ and
$-0.78^{+0.037}_{-0.043}$ for the best fit of viscosity parameters
corresponding to the two limiting conditions respectively. This
value is comparatively higher than that predicted by the joint
analysis of WMAP+BAO+H0+SN data, which is around -0.93
\cite{komatsu1,chimanto1}.

From the expression for matter density, it is clear that it diverges
as the scale factor tends to zero, which indicates the existence of
Big-Bang at the origin. This is further confirmed by obtaining the
curvature scalar which also becomes infinity at the origin.

The evolution of the total bulk viscous parameter is studied for the
best estimates of the bulk viscous parameter corresponding to
equations (\ref{condition1}) and (\ref{condition2}). In the initial
epoch of expansion, the total bulk viscosity is found to be negative
and hence violating the local second law of thermodynamics. But it
become positive from $z\leq 0.8$, from there onwards the local
second law is satisfied. However we found that the generalized
second law is satisfied throughout the evolution of the universe.

Since the model predicts the late acceleration of the universe as
like the standard forms of dark energy, we have analyzed the model
using statefinder parameters to distinguish it from other standard
dark energy models especially from $\Lambda$CDM model. The evolution
of the present model in the $\{r,s\}$ plane is shown in figure
\ref{fig:statefinder1} and it shows that the evolution of the
\{r,s\} parameter is in such a way that $r>1,s<0$, a feature similar
to the Chaplygin gas model. The present position of the bulk viscous
model in the r-s plane corresponds to $\{r_0,s_0\} =
\{1.25,-0.07\}$. Hence the model is distinguishably different from
the $\Lambda$CDM model.

Even though the model predicts the late acceleration, it failed
particularly in predicting the age of the universe and equation of
state parameter. It also fails with regard to the validity of the
local second law of thermodynamics even though the generalized
second law is satisfied. A similar model was studied in reference
\cite{av3}, where the authors have ruled out the possibility of bulk
viscous dark matter as a candidate of dark energy. But their
analysis is essentially a two parameter one since they took either
$\tilde{\zeta}_{1}$ or $\tilde{\zeta}_{2}$ as zero with
$\tilde{\zeta}_{0}>0.$ In the present work we have evaluated
$\tilde{\zeta}_{0}$, $\tilde{\zeta}_{1}$ and $\tilde{\zeta}_{2}$
simultaneously and found that there is a possibility for
$\tilde{\zeta}_{0}<0$ which gives a similar evolution of the
cosmological parameters as with $\tilde{\zeta}_{0}>0$. A crucial
test of this model is whether it predict the conventional radiation
dominated phase in the early universe. For this, one has to study
the phase space structure of this model and that will be a subject
of our future study. Such a study may also remove the apparent
degeneracy in the best estimated values of the bulk viscous
parameters.

In reference \cite{li1}, the authors have considered a unified model
for the dark sectors with a single component universe consisting of
bulk viscous dark matter, with the viscosity coefficient as a
function of density alone. They have found that in the background
level the model predicts an early deceleration and a late
acceleration. They also have analyzed the evolution of the first
order density perturbation. Regarding the density perturbation
growth, the authors have shown that for $\zeta(\rho)=\alpha \rho
^{m}$, with $m=-0.4$ and $\alpha \propto 0.236$, the density
perturbations behave drastically different from that of cold dark
matter in such a way that the presence of the viscosity becomes
significant and rapidly damped out the density perturbations at
small scales. This also causes the decay of gravitational potential
and hence modifies the large scale CMB spectrum. The authors have
pointed out that if $\zeta$ becomes a function of $H$ and $\dot{H}$,
like our case, the situation becomes more complex and would enhance
the damping of the perturbation growth. A similar study was also
carried out in reference \cite{velten1}. In this, the authors have
considered the ansatz $\zeta \propto \rho^{\nu}$ for the coefficient
of bulk viscosity and with $\nu = \frac{1}{2}$, the model mimics the
$\Lambda$CDM background evolution. They have shown that the viscous
dark fluids contribute to ISW Effect and thereby suppressing the
structure growth at small scales.

An important effect with which the model is to be contrasted is the
Integrated Sachs-Wolfe effect(ISW). The ISW Effect is the change in
the energy of a CMB photon as it passes through the evolving
gravitational potential wells \cite{Sachs}. For large time, the
behaviour of $a$ tends to that of the $\Lambda$CDM model for which
$\phi \sim 1+z$. So compared to the time of decoupling
($z\sim1090$), the potential will be diluted at later times which
consequentially causes the ISW effect. In the appendix below, we
have presented a brief argument regarding the ISW effect in the present model.\\[.2 in]

{\bf Acknowledgment}\\
The authors are thankful to the referee for the critical comments
which helped to improve the manuscript. The authors are also
thankful to Prof. Sahni for the valuable discussions. One of the
authors (TKM) is thankful to IUCAA, Pune for the hospitality, where
part of the work has been carried out. The author (AS) is
thankful to DST for giving financial support through INSPIRE fellowship. \\[.2 in]

\noindent{\bf Appendix}\\[0.1 in]
ISW effect \\[0.05 in]
Viscous dark matter will, in general, resist to the density
perturbations. Consequently it will dilute the gravitational
potential at the perturbed regions. This will subsequently affect
the CMB radiation and leads to ISW effect.

The ISW Effect is the change in the energy of a CMB photon as it
passes through the evolving gravitational potential wells. It is
obtained as
\begin{equation}
(\frac{\Delta
T}{T})_{ISW}=2\int^{\eta_{0}}_{\eta_{r}}\Phi'[(\eta_{0}-\eta)\bf{\hat{n}},\eta]d\eta
\end{equation}
where $\hat{n}$ is the photon trajectory and $\eta_{0}$ is the
conformal time today and $\eta_{r}$ is the conformal time at
recombination, $\Phi$ is the gravitational potential and prime
represents derivative with respect to the conformal time.

So the first step towards the calculation of the ISW Effect is to
obtain the evolution of gravitational potential in an expanding
universe. This can be obtained from Einstein's equation by taking
care of the perturbations. Viscous dark matter may cause a fast
decay of gravitational potential which modifies the CMB spectrum.

In Fourier space, the gravitational potential takes the form
\cite{Cooray}
\begin{equation}
\Phi=\frac{3}{2}\frac{\Omega_{m
o}}{a}(\frac{H_{0}}{k})^{2}\delta(k,\eta)
\end{equation}
where density perturbation, $\delta(k,\eta)=G(\eta)\delta(k,0)$.
$G(\eta)$ is the growth factor which is related to the Hubble
parameter as,

\begin{equation}
G(\eta)\propto\frac{H(\eta)}{H_0}\int_{z(\eta)}^\infty
dz'(1+z')(\frac{H_0}{H(z')})^3
\end{equation}

In matter dominated universe, $G\propto a$, so $\Phi$ remains a
constant, hence no ISW effect.

In our model, by considering the bulk viscous coefficient
$\zeta=\zeta_{0}+\zeta_{1}\frac{\dot{a}}{a}+\zeta_{2}\frac{\ddot{a}}{\dot{a}}$,
the Hubble parameter evolves as equation (\ref{hubbleina}). By using
this relation, the integral in the growth factor becomes
hypergeometric function. For simplification, let us consider the
case when $a$ is large, then $H\propto
a^{\frac{\tilde{\zeta}_{1}+\tilde{\zeta}_{2}-3}{2-\tilde{\zeta}_{2}}}$.
Then the growth factor becomes,
\begin{equation}
G \propto
(1+z)^{\frac{\tilde{\zeta}_{1}+\tilde{\zeta}_{2}-3}{2-\tilde{\zeta}_{2}}}\left(\frac{(2-\tilde{\zeta}_{2})
z^{\frac{-3\tilde{\zeta}_{1}+\tilde{\zeta}_{2}+5}{2-\tilde{\zeta}_{2}}}}{-3\tilde{\zeta}_{1}-\tilde{\zeta}_{2}+5}\right)
\end{equation}
So, potential becomes $\Phi \propto z^{8.34} (1 + z)^{4.45}$ (by
using extracted parameter values). From the last scattering surface,
which corresponds to $z=1091$, to the present epoch $z=0$, the
potential will be rarefied. This causes ISW effect. However, only
with an exact calculations and by obtaining the correlation
function, one can get the total ISW effect and its effect on the
structure formation.



\begin{thebibliography}{99}
\bibitem{Riess1} A. G. Riess et al., Astron. J. {\bf 116}, 1009 (1998)
\bibitem{Perl1} S. Perlmutter et al., Astrophys. J. {\bf 517} 565 (1999)
\bibitem{Bennet1} C. L. Bennet et al., Astrophys. J. Suppl. {\bf 148} 1 (2003)
\bibitem{Tegmark1} M. Tegmark et al., Phys. Rev. D {\bf 69} 103501 (2004)
\bibitem{weinberg1} S. Weinberg, Rev. Mod. Phys. {\bf 61} 1 (1989)
\bibitem{cope} E. J. Copeland, M. Sami and S. Tsujikawa, Int. J. Mod. Phys.
D {\bf 15} 1753 {2006}
\bibitem{fujii} Y. Fujii, Phys. Rev. D {\bf 26} 2580 (1982)
\bibitem{carroll} S. M. Carroll, Phys. Rev. Lett. {\bf 81} 3067 (1998)
\bibitem{chiba1} T. Chiba,T. Okabe and M. Yamaguchi, Phys. Rev. D {\bf 62} 023511 (2000)
\bibitem{kamen1} A. Y. Kamenshchik, U. Moschella, and V. Pasquier, Phys.
Lett. B {\bf 511} 265 (2001)
\bibitem{capo1} S. Capozziello, Int. J. Mod. Phys. D {\bf 11}
483 (2002)
\bibitem{ferraro1} R. Ferraro and F. Fiorini, Phys. Rev. D {\bf 75}
084031 (2007)
\bibitem{nojiri} S. Nojiri, S. D. Odintsov and M. Sasaki, Phys. Rev. D {\bf 71}
123509 (2005)
\bibitem{pad2} T. Padmanabhan and D. Kothawala, Phys. Rept. {\bf 531} 115 (2013)
\bibitem{horava1} P. Horava, Phys. Rev. D {\bf 79}
084008 (2009)
\bibitem{amendola1} L. Amendola, Phys. Rev. D {\bf 60}
043501 (1999)
\bibitem{dvali1} G. R. Dvali, G. Gabadadze, and M. Porrati, Phys. Lett. B {\bf 485}
208 (2000)
\bibitem{pad} T. Padmanabhan and S. M. Chitre, Phys. Lett. A {\bf 120}
443 (1987)
\bibitem{waga} I. Waga, R. C. Falcao and  R. Chanda, Phys. Rev. D {\bf 33}
1839 (1986)
\bibitem{fabris1} J.C. Fabris, S.V.B. Goncalves and R. de S\'{a} Ribeiro,
Gen. Rel. Grav. {\bf 38} 495 (2006)
\bibitem{li1} B. Li and J. D. Barrow,
Phys. Rev D {\bf 79} 103521 (2009)
\bibitem{Hiplito1} W. S. Hip\'{o}lito-Ricaldi, H. E. S. Velten and
W. Zimdahl, Phys Rev D {\bf 82} 063507 (2010).
\bibitem{av1} A. Avelino and U. Nucamendi, JCAP {\bf 04}
006 (2009)
\bibitem{av2} A. Avelino and U. Nucamendi, JCAP {\bf 08} 009 (2010)
\bibitem{Zimdahl1} W. Zindahl, D. J. Schwarz,  A. B. Balakin and D. Pav\'{o}n
, Phys. Rev D {\bf 64} 063501 (2001)
\bibitem{wilson1} J. R. Wilson, G. J. Mathews and G. M. Fuller, Phys.
Rev. D {\bf 75} 043521 (2007)
\bibitem{okumura1} H. okumura and
 F. Yonezawa, Physica A {\bf 321}
 207 (2003)
\bibitem{ilg1} P. Ilg and H. C. Ottinger, Phys. Rev. D {\bf 61}
023510 (2000)
\bibitem{av3} A. Avelino et al., JCAP {\bf 8} 012 (2013)
\bibitem{weinberg2} S. Weinberg, \emph{Gravitation and
cosmology: principles and applications of the general theory of
relativity}, John Wiley \& sons Inc., New york U.S.A. (1972).
\bibitem{misner1} C. W. Misner, K. S. Thorne and J. A. Wheeler,
\emph{Gravitation}, W.H. Freemann and Company, U.S.A. (1973).
\bibitem{Eckart1} C. Eckart, Phys. Rev. {\bf 58} 919 (1940)
\bibitem{Landau1} L. D. Landau and E. M. Lifshitz, \emph{Fluid Mechanics} Addison-Wesley, Reading U.S.A.
(1959).
\bibitem{Israel1} W. Israel, Ann. Phys. (N.Y.) {\bf 100}
310 (1976)
\bibitem{Hiscock1} W. A. Hiscock and L. Lindblom, Phys. Rev.
D {\bf 31} 725 (1985)
\bibitem{Israel2} W. Israel and J. M. Stewart, Ann. Phys. (N.Y.) {\bf 118}
341 (1979)
\bibitem{Israel3} W. Israel and J. M. Stewart, Proc. R. Soc. Lond. A {\bf 365} 43 (1979)
\bibitem{kremer1} G. M. Kremer and F. P. Devecchi, Phys. Rev.
D {\bf 67} 047301 (2003)
\bibitem{cataldo1}  M. Cataldo, N. Cruz and S. Lepe, Phys. Lett.
B {\bf 619} 5 (2005)
\bibitem{hu1} M. G. Hu and X. H. Meng, Phys. Lett. B {\bf 635}
186 (2006)
\bibitem{hiscock1} W. A. Hiscock and J. Salmonson, Phys. Rev. D {\bf 43}
3249 (1991)
\bibitem{Zimdahl} W. Zimdahl, Phys. Rev. D, {\bf 53} 5483 (1996)
\bibitem{Zakari} M. Zakari and D. Jou, Phys. Rev. D, {\bf 48} 1597 (1993)
\bibitem{pavon1} D. Pav\'{o}n, D. Jou and J. Casas-V\'{a}zquez,
Ann. Inst. Henri Poincar\'{e} {\bf 36} 79 (1982)
\bibitem{ren1} J. Ren and Xin-He Meng, Phys. Lett. B {\bf 633} 1 (2006)
\bibitem{tegmark2} M. Tegmark et al., Phys. Rev. D {\bf 74}
123507 (2006)
\bibitem{carretta1} E. Carretta, R. G. Gratton, G. Clementini and F. Fusi Pecci, AstroPhys. J. {\bf533} 215 (2000)
\bibitem{wmap1} WMAP Colabration,G. Hinshaw et. al., AstroPhys. J. Suppl. {\bf 180}
225 (2009)
\bibitem{praseetha1} P. Praseetha and T. K. Mathew,
Int. J. Mod. Phys. D {\bf 23} 1450024 (2014)
\bibitem{alcaniz1} J. S. Alcaniz and J. A. S. Lima, Astrophs. J. {\bf 521}
L87 (1999)
\bibitem{Brevik2} I. Brevik, E. Elizalde, S. Nojiri and S.D.
Odintsov, Phys.Rev. D {\bf 84} 103508 (2011)
\bibitem{komatsu1} E. Komatsu et. al, WMAP Colabration, AstroPhys. J. Suppl. {\bf 192}
18 (2011)
\bibitem{chimanto1} L.P. Chimanto and M. G. Richarte, Phys. Rev. D {\bf 84} 123507 (2011)
\bibitem{alam1} U. Alam, V. Sahni and A. A. Starobinsky, JCAP {\bf 0406}
008 (2004)
\bibitem{Brevik} I. Brevik O. Gr{\o}n, Astrophys. Space Sci. {\bf 347} 399 (2013)
\bibitem{gibbons1} G. W. Gibbons, S. W. Hawking, Phys. Rev. D {\bf 15} 2738 (1997)
\bibitem{tkm1} T. K. Mathew, R. Aiswarya and K. S. Vidya, Eur. Phys. J. C {\bf 73} 2619 (2013)
\bibitem{Sheykhi1} A. Sheykhi, Class.
Quantum Grav. {\bf 27} 025007 (2010)
\bibitem{setare1} M. R. Setare and A. Sheykhi, Int. J. Mod. Phys. D {\bf 19}
1205 (2010)
\bibitem{davis1} P. C. W. Davis, Class. Quantum Grav. {\bf 4}
L225 (1987)
\bibitem{liu} D. J. Liu and W. Z. Liu, Phys. Rev. D {\bf 77}
027301 (2008)
\bibitem{kowalski1} M. Kowalski et al., Astrophys. J. {\bf 686} 749 (2008)
\bibitem{sahni1} V. Sahni, T. D. Saini, A. A. Starobinsky, and U. Alam, JETP Lett. {\bf77}
201 (2003)
\bibitem{wu1} B. Wu Ya, S. Li, M. H. Fu and J. He, Gen. Relativ. Gravity {\bf39}
653 (2007)
\bibitem{Sachs} R.K. Sachs and A.M. Wolfe, Astrophys. J. {\bf 147} 73 (1967)
\bibitem{velten1} Hermano Velten and Dominik J. Schwarz, JCAP {\bf 09} 016 (2011)
\bibitem{Cooray} Asantha Cooray, Phys. Rev. D {\bf 65} 103510 (2002)

\end{thebibliography}
\end{document}